\documentclass[superscriptaddress,aps,prx,twocolumn,showpacs,nofootinbib, longbibliography, notitlepage,floatfix,]{revtex4-2}

\usepackage{amsmath,amssymb,amsthm}

\usepackage[pdftex]{graphicx}
\usepackage{times}
\usepackage{braket}
\usepackage{comment}
\usepackage{physics}
\usepackage{color}
\usepackage{natbib}
\usepackage[dvipsnames]{xcolor}
\usepackage{float}
\makeatletter
\let\newfloat\newfloat@ltx
\makeatother
\usepackage{algorithm}
\usepackage{algpseudocode}
\usepackage{mathtools}
\usepackage[normalem]{ulem}
\usepackage{latexsym}
\usepackage{tabularx, booktabs}
\usepackage{graphics,epstopdf}
\usepackage{graphicx}
\usepackage{float}
\usepackage{amsfonts}
\usepackage{tikz}
\usetikzlibrary{quantikz}
\usepackage{color,soul}

\newcommand{\be}{\begin{equation}}
\newcommand{\ee}{\end{equation}}
\newcommand{\ba}{\begin{eqnarray}}
\newcommand{\ea}{\end{eqnarray}}

\usepackage{multirow}
\usepackage{appendix}
\usepackage{url}
\renewcommand{\arraystretch}{1.7}

\newcommand{\eqlabel}[1]{Eq.~\eqref{#1}}
\newcommand{\figlabel}[1]{Fig.~\ref{#1}}
\newcommand{\applabel}[1]{Appendix~\ref{#1}}
\newcommand{\seclabel}[1]{Sec.~\ref{#1}}

\usepackage[colorlinks=true,citecolor=blue,urlcolor=blue]{hyperref}

\begin{document}
\title{Digitized Counterdiabatic Quantum Sampling} 

\author{Narendra N. Hegade}
\email{narendra.hegade@kipu-quantum.com}
\affiliation{Kipu Quantum GmbH, Greifswalderstrasse 212, 10405 Berlin, Germany}

\author{Nachiket L. Kortikar}
\affiliation{Kipu Quantum GmbH, Greifswalderstrasse 212, 10405 Berlin, Germany}

\author{Balaganchi A. Bhargava}
\affiliation{Kipu Quantum GmbH, Greifswalderstrasse 212, 10405 Berlin, Germany}

\author{Juan F. R. Hernández}
\affiliation{Kipu Quantum GmbH, Greifswalderstrasse 212, 10405 Berlin, Germany}

\author{Alejandro Gomez Cadavid}
\affiliation{Kipu Quantum GmbH, Greifswalderstrasse 212, 10405 Berlin, Germany}
\affiliation{Department of Physical Chemistry, University of the Basque Country EHU, Apartado 644, 48080 Bilbao, Spain}

\author{Pranav Chandarana}
\affiliation{Kipu Quantum GmbH, Greifswalderstrasse 212, 10405 Berlin, Germany}
\affiliation{Department of Physical Chemistry, University of the Basque Country EHU, Apartado 644, 48080 Bilbao, Spain}

\author{Sebastián V. Romero}
\affiliation{Kipu Quantum GmbH, Greifswalderstrasse 212, 10405 Berlin, Germany}
\affiliation{Department of Physical Chemistry, University of the Basque Country EHU, Apartado 644, 48080 Bilbao, Spain}

\author{Shubham Kumar}
\affiliation{Kipu Quantum GmbH, Greifswalderstrasse 212, 10405 Berlin, Germany}

\author{Anton Simen}
\affiliation{Kipu Quantum GmbH, Greifswalderstrasse 212, 10405 Berlin, Germany}
\affiliation{Department of Physical Chemistry, University of the Basque Country EHU, Apartado 644, 48080 Bilbao, Spain}

\author{Anne-Maria Visuri}
\affiliation{Kipu Quantum GmbH, Greifswalderstrasse 212, 10405 Berlin, Germany}

\author{Enrique Solano}
\email{enr.solano@gmail.com}
\affiliation{Kipu Quantum GmbH, Greifswalderstrasse 212, 10405 Berlin, Germany}

\author{Paolo A. Erdman}
\email{paolo.erdman@kipu-quantum.com}
\affiliation{Kipu Quantum GmbH, Greifswalderstrasse 212, 10405 Berlin, Germany}

\date{\today}
\begin{abstract}
We propose digitized counterdiabatic quantum sampling (DCQS), a hybrid quantum-classical algorithm for efficient sampling from energy-based models, such as low-temperature Boltzmann distributions. The method utilizes counterdiabatic protocols, which suppress non-adiabatic transitions, with an iterative bias-field procedure that progressively steers the sampling toward low-energy regions. We observe that the samples obtained at each iteration correspond to approximate Boltzmann distributions at effective temperatures. By aggregating these samples and applying classical reweighting, the method reconstructs the Boltzmann distribution at a desired temperature. We define a scalable performance metric, based on the Kullback–Leibler divergence and the total variation distance, to quantify convergence toward the exact Boltzmann distribution. DCQS is validated on one-dimensional Ising models with random couplings up to $124$ qubits, where exact results are available through transfer-matrix methods. We then apply it to a higher-order spin-glass Hamiltonian with $156$ qubits executed on IBM quantum processors. We show that classical sampling algorithms, including Metropolis-Hastings and the state-of-the-art low-temperature technique parallel tempering, require up to three orders of magnitude more samples to match the quality of DCQS, corresponding to an approximately $2\times$ runtime advantage. Boltzmann sampling underlies applications ranging from statistical physics to machine learning, yet classical algorithms exhibit exponentially slow convergence at low temperatures. Our results thus demonstrate a robust route toward scalable and efficient Boltzmann sampling on current quantum processors.
\end{abstract}

\maketitle

\section{Introduction}
Sampling from complex probability distributions such as Boltzmann distributions is a fundamental challenge across various disciplines, including statistical physics \cite{huang1988}, molecular dynamics \cite{frenkel2002,noe2019}, machine learning \cite{ackley1985} and finance \cite{glasserman2003}. 
Traditional sampling methods based on Markov chain Monte Carlo (MCMC) \cite{dongarra2000} are widely used but often struggle with high-dimensional distributions and in the low-temperature regime, due to slow convergence and poor mixing. As illustrated in~\figlabel{fig:sketch}(a), as the temperature tends to zero, the autocorrelation (mixing) time of the Metropolis-Hastings (MH) MCMC method with local updates diverges \cite{newman1999}, making it impractical in this regime. Furthermore, in the strict zero-temperature limit, sampling from the Boltzmann distribution reduces to finding the ground state, which is, in the worst-case scenario, an NP-hard problem, as for spin glasses \cite{barahona1982}.

The classical state-of-the-art technique for low-temperature sampling, sketched in \figlabel{fig:sketch}(b), is parallel tempering (PT) \cite{hukushima1996, earl2005}, also known as replica exchange. It mitigates slow mixing by running multiple MCMC simulations in parallel at different temperatures and periodically attempting exchanges between them. While this can prevent the diverging autocorrelation time and allows escaping local minima, it comes at the expense of running multiple auxiliary simulations.

Recent years have seen growing interest in employing
quantum computers to perform Boltzmann sampling, taking advantage of quantum rather than thermal fluctuations. Given their open nature, quantum annealers have been proposed as noisy Boltzmann samplers \cite{vuffray2022}. To improve their accuracy,
various adjustments have been proposed, such as introducing spurious hardware-dependent couplings \cite{vuffray2022}, limiting the class of simulable systems \cite{nelson2022}, focusing on the low-temperature regime \cite{sandt2023}, restricting the validity to the high-temperature regime \cite{gyhm2024}, and using embedding schemes \cite{teza2025}.

\begin{figure*}[!tb]
    \centering
     \includegraphics[width=\textwidth]{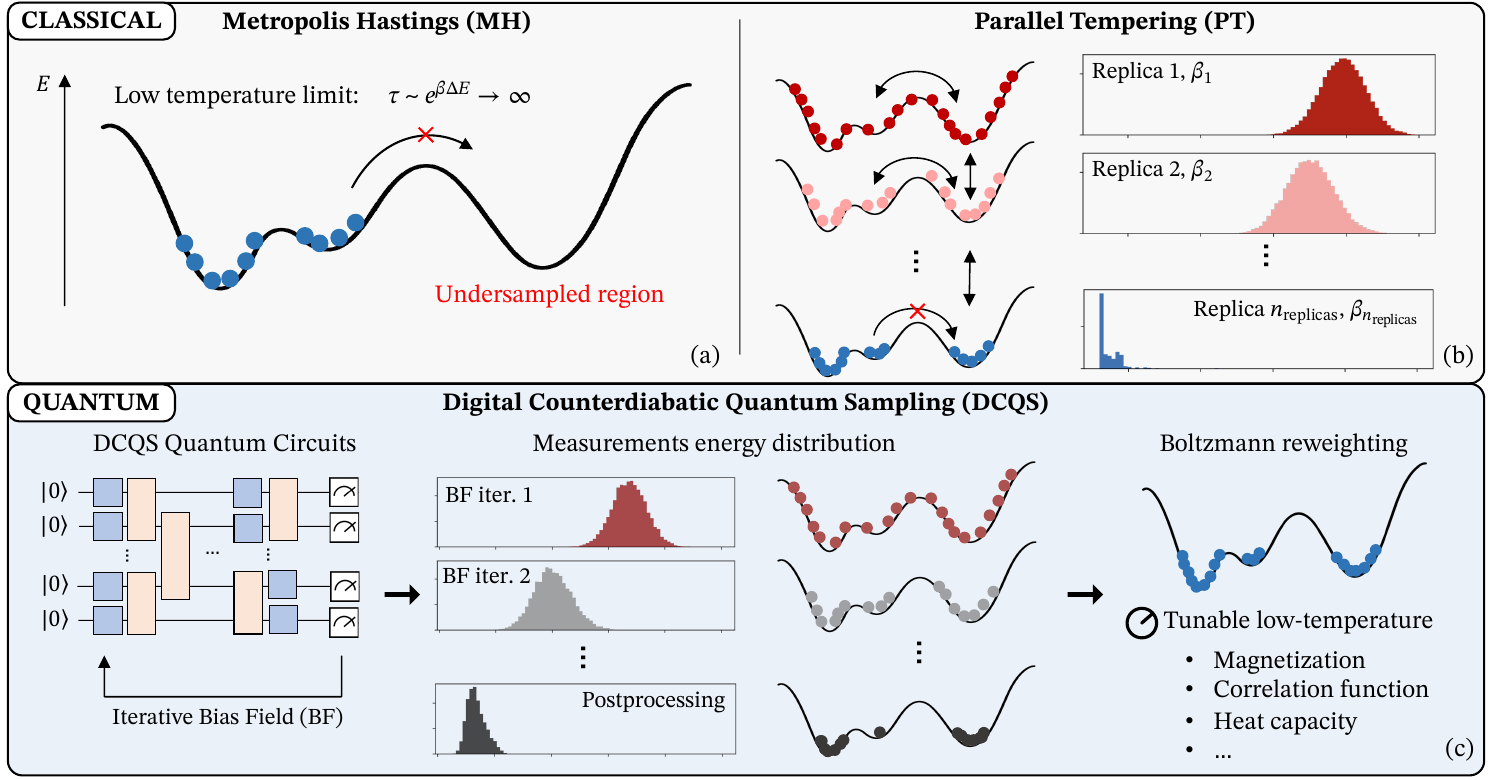}
    \caption{Sketch of (a,b) classical methods for Boltzmann sampling and (c) digitized counterdiabatic quantum sampling (DCQS). (a) The classical Metropolis-Hasting algorithm suffers from a diverging mixing time at low temperatures, since it can get trapped in local minima. The black line pictorially represents the energy landscape of the classical Hamiltonian. Transitions to the local minima on the right are exponentially suppressed in the low-temperature limit. (b) Parallel tempering (PT) is a classical method that overcomes this limitation by running multiple ``replicas'' in parallel (one per row), each one at a different inverse temperature $\beta_i$. The left column represents the allowed transitions within the energy landscape (horizontal arrows), and between adjacent replicas (vertical arrows), while the right column shows the corresponding energy distributions. This leads to a better low-temperature sampling at the expense of an increased simulation cost. (c) DCQS relies on running quantum circuits with an iterative bias field strategy on a digital quantum computer (left), producing lower and lower energy states (center), which are then reweighed in order to reproduce low-temperature observables (right). }
    \label{fig:sketch}
\end{figure*}
Research on employing digital quantum computers as Boltzmann samplers remains limited and requires further conceptual advances, as the unitary evolution of closed quantum systems is, at first glance, not directly connected to the Boltzmann distribution that characterizes open-system statistics.
Variational quantum imaginary time evolution has been proposed to train Boltzmann machines, and has been tested on an IBM quantum processing unit (QPU) \cite{ibm} with up to $20$ qubits~\cite{zoufal2021}. Specifically designed Hamiltonians, whose ground-state measurement outcomes mimic the Boltzmann distribution, have been proposed~\cite{wild2021}, as well as sampling states using quantum alternating operator ansatz~\cite{pelofske2025}. However, the implementation of these approaches on QPUs has not been investigated.
Alternatively, quantum sub-routines have been proposed to improve the quality of classical MCMC methods. For instance, the autocorrelation time of MCMC has been decreased using either quantum annealers \cite{scriva2023} or digital quantum computers \cite{layden2023} during the proposal step.

We introduce digitized counterdiabatic quantum sampling (DCQS) to overcome the difficulty of classical methods in sampling Boltzmann distributions at low temperatures, the lack of scalable quantum algorithms suitable for digital noisy intermediate-scale quantum (NISQ) computers, and the scarcity of large-scale (100+ qubits) experimental validations. It is a NISQ-compatible quantum-classical algorithm based on counterdiabatic protocols \cite{demirplak2003, berry2009, delcampo2013}, designed to efficiently sample low-temperature Boltzmann distributions and compute thermodynamic observables of classical Hamiltonians using digital quantum computers.

The proposed method, illustrated in  \figlabel{fig:sketch}(c), relies on the observation that, in the limit of infinitely slow driving, adiabatic quantum evolution preserves the system in its instantaneous ground state, thus yielding the zero-temperature limit of the Boltzmann distribution of the final Hamiltonian~\cite{gyhm2024}.
In practice, however, coherent transitions due to non-adiabatic driving \cite{landau1932,zener1932}, noise in NISQ devices, discretization, and Trotter errors \cite{lloyd1996} lead to excitations in the final state~\cite{visuri2025}. We utilize these apparently undesired effects to efficiently obtain not only the ground state but also low-energy states that predominantly contribute to the low-temperature Boltzmann distribution. 
To turn this idea into a quantum algorithm compatible with NISQ devices, we employ counterdiabatic protocols with an adaptive bias field \cite{cadavid2025,romero2025,iskay} to mimic the adiabatic behavior in shorter times, leading to low-depth quantum circuits that run within the device's coherence time. Finally, we employ post-processing techniques and a reweighting scheme \cite{sandt2023} to compute low-temperature expectation values. 

We first validate our approach by computing thermal observables for the 1D Ising model with random coefficients. We compare both exact results and classical simulations to DCQS executed on the \textsc{IBM Marrakesh} QPU \cite{ibm}. 
We then address a more challenging three-local Hamiltonian with $156$ spins and hundreds of Pauli terms, showing that DCQS executed on \textsc{IBM Fez} outperforms both MCMC and PT in the low-temperature regime in terms of sampling efficiency: PT requires three orders of magnitude more samples to match the quality of the DCQS distribution, corresponding to a $\sim 2\times$ runtime advantage over an optimized single-core PT implementation.
For this comparison, we introduce a figure of merit based on the Kullback–Leibler divergence \cite{kullback1951} and the total variation distance \cite{lecam1973}, which quantifies the distance between approximate and exact Boltzmann distributions. This figure of merit is scalable to large qubit numbers and also permits convergence assessment in the absence of a known ground truth.

The manuscript is organized as follows. The Boltzmann sampling problem and methodology are introduced in Sec.~\ref{sec:boltzmann}, where we introduce the DCQS algorithm in Sec.~\ref{subsec:dcqs}. We provide an overview of the MH and PT classical sampling methods in Sec.~\ref{subsec:classical_sampling}, and we introduce the scalable figure of merit in Sec.~\ref{subsec:logz}. We present our results in Sec.~\ref{sec:results}: in Sec.~\ref{subsec:ising}, we benchmark DCQS, tackling the 1D Ising model with random coefficients up to $124$ qubits, and in Sec.~\ref{subsec:3body}, we apply it to a three-body Hamiltonian with $156$ qubits and compare its performance against classical sampling methods. In Sec.~\ref{sec:conclusions}, we draw the conclusions and present future outlooks.

\section{Boltzmann Sampling Problem}
\label{sec:boltzmann}
Let us consider an $N$-qubit quantum computer and a classical Hamiltonian $\hat{H}_\text{f}$ that is diagonal in the computational basis. Let $\mathcal{S}$ denote the set of all possible $2^N$ bitstrings $s$, each representing an element of the computational basis $\{\ket{s}\}$. The corresponding energies $E(s)$ are defined by $\hat{H}_\text{f}\ket{s} = E(s)\ket{s}$. 
The Boltzmann distribution  $\mu(s)$, associated with the classical Hamiltonian $\hat{H}_\text{f}$, is given by
\begin{equation}
    \mu(s) = \frac{e^{-\beta E(s)}}{\mathcal{Z}}, \quad \mathcal{Z} = \sum_{s\in \mathcal{S}} e^{-\beta E(s)},
    \label{eq:boltzmann_def}
\end{equation}
where $\beta = 1/(k_BT)$ is the inverse temperature, $k_B$ is Boltzmann's constant, and $\mathcal{Z}$ is the partition function \cite{huang1988}. In the following, we use natural units where $k_B = \hbar = 1$.
Our goal is to develop a quantum algorithm to compute the thermal expectation value of an arbitrary observable $\hat{O}$, given by
\begin{equation}
    \ev*{\hat{O}} =     \sum_{s\in\mathcal{S}} \mu(s) \, O(s),
    \label{eq:o_avg}
\end{equation}
where $O(s) = \mel*{s}{\hat{O}}{s}$ is the expectation value of the observable in state~$s$. 

\subsection{Digitized Counterdiabatic Quantum Sampling}
\label{subsec:dcqs}
\subsubsection{Sampling method}
In this section, we describe the DCQS algorithm, which aims to efficiently compute thermal observables of many-body systems in the low-temperature regime, where classical approaches struggle with diverging mixing time.
We start by noting that, because of the exponential energy dependence of $\mu(s)$, only a small fraction of low-energy states, among all possible states $s\in \mathcal{S}$, significantly contribute to the Boltzmann distribution at sufficiently low temperatures \cite{sandt2023}. 
We therefore define the low-temperature regime as that in which the number of states that give a finite contribution to the calculated observables [see \eqlabel{eq:o_avg}] is at most polynomial in $N$.
This assumption holds whenever the density of states of the classical Hamiltonian $\hat{H}_\text{f}$ does not increase exponentially with energy, which is the case for most non-pathological classical Hamiltonians $\hat{H}_\text{f}$. A practical way to evaluate whether a given set of states is sufficient to accurately describe the Boltzmann distribution will be presented in Sec.~\ref{subsec:logz}.

We employ a hybrid quantum-classical approach: first, a quantum algorithm is used to determine the relevant low-energy states. Then, a classical reweighting scheme is used to compute thermal expectation values of arbitrary observables. 
Let us denote with $\tilde{\mathcal{S}}\subseteq \mathcal{S}$ the set of states determined using the quantum algorithm described below, which should correspond to the relevant low-energy states. We can compute any observable by replacing the summation in \eqlabel{eq:o_avg} with
\begin{equation}
    \ev*{O} = \sum_{\tilde{s}\in\tilde{\mathcal{S}}} \tilde{\mu}(\tilde{s}) \, O(\tilde{s}),
    \label{eq:o_tilde_avg}
\end{equation}
where
\begin{equation}
    \tilde{\mu}(\tilde{s}) = \frac{1}{\tilde{\mathcal{Z}}} \cdot
    \begin{cases}
        e^{-\beta E(\tilde{s})} & \text{if } \tilde{s}\in \tilde{\mathcal{S}}, \\
        0 & \text{otherwise},
    \end{cases}
    \quad
    \tilde{\mathcal{Z}} = \sum_{\tilde{s}\in\tilde{\mathcal{S}}} e^{-\beta E(\tilde{s})},
    \label{eq:tilde_p_def}
\end{equation}
$\tilde{\mathcal{Z}}$ being the approximate partition function. Notice that while \eqlabel{eq:o_avg} is a summation over $2^N$ bitstrings, \eqlabel{eq:o_tilde_avg} only sums over the small subset of states $\tilde{\mathcal{S}}$ that scales polynomially with $N$ in the low-temperature regime; thus, it can be performed efficiently on a classical computer.

We now discuss the quantum algorithm used to produce the relevant low-energy states $\tilde{S}$.
The method is inspired by the quantum adiabatic theorem \cite{sakurai2017}. Let us consider the following time-dependent Hamiltonian, 
\begin{equation}
    \hat{H}_\text{ad}(\lambda_t) = (1-\lambda_t)\,\hat{H}_\text{i} + \lambda_t\,\hat{H}_\text{f},
    \label{eq:h_ad}
\end{equation}
 where $\hat{H}_\text{i}$ is an initial Hamiltonian, $\lambda_t$ is a time-dependent control satisfying $\lambda_0=0$ and $\lambda_\tau=1$, and $\tau$ is the evolution time.
 The adiabatic theorem states that a system initialized in the ground state of $\hat{H}_\text{i}$ and evolved unitarily according to the time-dependent Hamiltonian $\hat{H}_\text{ad}(\lambda_t)$ in \eqlabel{eq:h_ad} will end in the ground state of $\hat{H}_\text{f}$, provided that $\tau$ is sufficiently long.
In finite time, non-adiabatic transitions can occur, so that the states measured at the final time may be excited states of $\hat{H}_\text{f}$. We aim to take advantage of this apparent detrimental effect and use it to identify the relevant low-energy states.
One approach would be to implement the time-evolution operator corresponding to \eqlabel{eq:h_ad} on a digital quantum computer. However, finding low-energy states requires a large $\tau$, thus a large number of Trotter steps, leading to deep quantum circuits that are not compatible with current NISQ devices. 

To reduce the circuit depth, we employ a counterdiabatic (CD) protocol \cite{demirplak2003, berry2009, delcampo2013}, which  allows us to mimic the adiabatic evolution in finite time by including an additional driving term $\dot{\lambda}_t\,\hat{A}_{\lambda_t}$ to the Hamiltonian,
\begin{equation}
    \hat{H}_\text{cd}(\lambda_t) = \hat{H}_\text{ad}(\lambda_t) + \dot{\lambda}_t\,\hat{A}_{\lambda_t}.
    \label{eq:h_ad_cd}
\end{equation}
Here, $\dot{\lambda}_t$ is the time-derivative of the control and $\hat{A}_{\lambda}$ is known as the adiabatic gauge potential \cite{sels2017}. In general, calculating the exact gauge potential requires diagonalizing $\hat{H}_\text{ad}(\lambda)$ \cite{berry2009}, which is computationally prohibitively expensive. Nonetheless, various proposals have been put forward to derive approximate gauge potentials \cite{hatomura2021, takahashi2024}. For instance, Ref.~\cite{claeys2019} expresses it as $\hat{A}_\lambda^{(l)} = i\sum_{k=1}^l\alpha_k(\lambda)\,\hat{\mathcal{O}}_{2k-1}(\lambda)$, where $l$ is the expansion order, and $\hat{\mathcal{O}}_k(\lambda) = [\hat{H}_\text{ad}(\lambda), \hat{\mathcal{O}}_{k-1}(\lambda)]$ with $\hat{\mathcal{O}}_0(\lambda) = \partial_\lambda \hat{H}_\text{ad}(\lambda)$. The $\alpha_k(\lambda)$ coefficients can be computed using a variational minimization approach (see \applabel{app:cd} for details).

To derive a quantum circuit for digital quantum computers, we use the first-order product formula \cite{barends2016}, with $n_\text{Trot}$ Trotter steps, to decompose the time-evolution operator $\hat{U}(t) = \overleftarrow{\mathcal{T}} \exp\left( -i \int_{0}^t \hat{H}_\text{cd}(\lambda_{t'}) \, dt' \right)$ induced by  \eqlabel{eq:h_ad_cd} \cite{hegade2021,hegade2022}, $\overleftarrow{\mathcal{T}}$ being the time-ordering operator. This leads to
\begin{equation}
    \hat{U}(\tau) \approx \prod_{k=1}^{n_\text{Trot}}\prod_{j=1}^{n_\text{terms}} \exp{ -i\,\Delta t\, \gamma_j(k\,\Delta t) \hat{P}_j },
    \label{eq:trott}
\end{equation}
where $\Delta t= \tau/n_\text{Trot}$ is the time-step, and $\hat{H}_\text{cd}(\lambda_t)=\sum_{j=1}^{n_\text{terms}} \gamma_j(t)\, \hat{P}_j$ is the decomposition of the CD Hamiltonian into the sum of $n_\text{terms}$ local Pauli operators $\hat{P}_j$.

We now discuss the choice of the initial Hamiltonian $\hat{H}_\text{i}$, whose ground state must be easy to prepare on a quantum computer. A typical choice in the context of adiabatic computation is $\hat{H}_\text{i} = -\sum_{i=1}^{N} \hat{X}_i$, where $\hat{X}_i, \hat{Y}_i, \hat{Z}_i$ denote the Pauli matrices acting on the $i$-th qubit. 
Here, to further enhance the probability of finding the relevant low-energy states, we employ a bias field (BF) in the initial Hamiltonian \cite{cadavid2025,romero2025,iskay}. More specifically, we consider an iterative state collection process divided into $n_\text{iter}$ iterations. 
 In the $k$-th iteration, with $k=1,\dots, n_\text{iter}$, we consider 
\begin{equation}
    \hat{H}_\text{i} = -\sum_{i=1}^{N} \left(\hat{X}_i + w \,b_i^{(k)} \hat{Z}_i\right)
\end{equation}
as initial Hamiltonian, where $b_i^{(k)}$ is the BF, and $w>0$ is the BF weight, which allows us to tune the relative strength of the BF with respect to the transverse field $-\sum \hat{X}_i$.
We then collect $n_\text{shots}$ bitstrings (samples), denoted with $\{s_j^{(k)}\}_{j=1}^{n_\text{shots}}$, ordered from the lowest to the highest energy. 
The value of the BF is set to zero during the first iteration, i.e. 
\begin{equation}
    b_i^{(1)}=0.
\end{equation}
For the following iterations, we compute the BF to steer the time-evolution towards the lowest-energy states that have been found in the previous iteration. In particular, we consider the $n_\text{cvar}$ lowest energy states and choose, for $k\geq 1$, 
\begin{equation}
    b_i^{(k+1)} = -\frac{1}{n_\text{cvar}}\sum_{j=1}^{n_\text{cvar}} \mel*{s_j^{(k)}}{\hat{Z}_i}{s_j^{(k)}}.
    \label{eq:bf_update}
\end{equation}
Note that $n_\text{iter}$, $w$, and $n_\text{cvar}$ are hyperparameters of DCQS that can be tuned to improve sampling efficiency. Interestingly, the different error sources, such as the Trotterization of the time-evolution operator, factorizing it into Pauli rotations, the use of an approximate adiabatic gauge potential, and potential hardware errors, 
tend to increase the probability of populating excited states rather than the ground state. This is not necessarily a detrimental effect, provided that we can still identify the relevant low-energy states with finite probability. Our algorithm is therefore inherently robust to a moderate amount of errors, which makes it suitable for NISQ devices.

\subsubsection{Approximate Boltzmann distributions from DCQS}
\begin{figure}[!tb]
    \centering
    \includegraphics[width=\columnwidth]{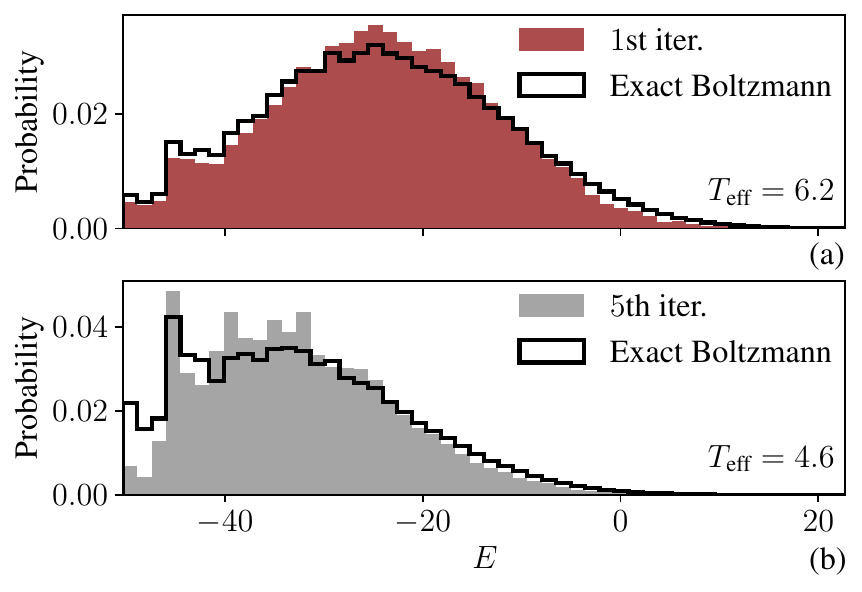}
    \caption{Energy distribution obtained from DCQS for the spin glass model in \eqlabel{eq:spin_glass} for $N=18$. The red distribution in panel (a) corresponds to the $1$st BF iteration with $b_i^{(1)}=0$, and the gray distribution in panel (b) to the $5$th iteration with iteratively updated $b_i^{(5)}$. The black thick line corresponds to the exact Boltzmann distribution with an effective temperature, reported on the plot, determined by matching the average energy of the DCQS distribution.
    The parameters are $n_\text{iter} =5$, $n_\text{shots} = 30\,000$, $w=1$, and $n_\text{cvar}=30\,000$.}
    \label{fig:spin_glass_energy_2}
\end{figure}
To further illustrate the intuition behind DCQS, we notice that while an ideal adiabatic evolution ($\tau\to\infty$) prepares the ground state of the final Hamiltonian, corresponding to the
zero-temperature limit of its Boltzmann distribution, the opposite limit of a sudden quench $\tau\to 0$ yields the ground state of the transverse-field term---an equal superposition of all bitstrings---whose measurement statistics corresponds to the infinite-temperature Boltzmann distribution. Therefore, we may speculate that the finite-time evolution, by digitized counterdiabatic evolution, may lead to a distribution resembling a finite-temperature Boltzmann distribution. To test this hypothesis empirically, we consider a spin-glass model
\begin{equation}
    \hat{H}_\text{f} = \left[
    \sum_{i=1}^{N}  h_i \, \hat{Z}_i + 
    \sum_{i<j}  J_{i,j} \, \hat{Z}_i \hat{Z}_j\right],
    \label{eq:spin_glass}
\end{equation}
where $h_i$ and $J_{i,j}$ are, respectively, the local fields and coupling constants that are sampled with uniform probability in the interval $[-E_0, E_0]$. In the following, energies, temperatures, and coupling constants are reported in units of the energy scale $E_0$.
We perform DCQS for $n_\text{iter}=5$ BF iterations and apply the approximations detailed in Sec.~\ref{sec:results}, which includes setting $n_\text{Trot}=2$. 
In \figlabel{fig:spin_glass_energy_2}, we present a comparison between the energy distribution obtained with DCQS, simulated using the exact state-vector simulator of IBM \cite{javadiabhari2024}, and the exact Boltzmann distribution whose temperature, reported on the figure as $T_\text{eff}$, is chosen such that the mean energy matches that of the DCQS distribution. This choice of the temperature is equivalent to minimizing the KL divergence between the DCQS distribution and the exact Boltzmann distribution (see \applabel{app:fit_temp} for details).
The red distribution in \figlabel{fig:spin_glass_energy_2}(a) corresponds to the energy of the bitstrings collected during the first BF iteration, i.e., without a BF, the gray distribution in \figlabel{fig:spin_glass_energy_2}(b) corresponds to the bitstrings obtained during the $5$th and last BF iteration, and the black thick line is the exact Boltzmann distribution. 

Despite the numerous approximations and the use of an approximate counterdiabatic driving, the energy distribution of DCQS and of the exact Boltzmann distribution in \figlabel{fig:spin_glass_energy_2} are remarkably similar, supporting our hypothesis that a finite-time evolution can produce an approximate finite-temperature Boltzmann distribution. In \applabel{app:dcqs_boltzmann}, we provide additional evidence by comparing the energy distributions and average magnetizations obtained experimentally on \textsc{IBM Kingston} for random quadratic Hamiltonians with $156$ qubits, with the corresponding fitted Boltzmann distribution computed using the Metropolis-Hastings algorithm (detailed in Sec.~\ref{subsec:classical_sampling}). Interestingly, comparing \figlabel{fig:spin_glass_energy_2}(a) with \figlabel{fig:spin_glass_energy_2}(b), we observe that the BF has the effect of lowering the effective temperature, which is intuitively expected, since it shifts the energy distribution towards lower energies. This feature is crucial for the success of DCQS at identifying low-energy states.
However, we also notice that there are slight deviations in the tails of the distribution in \figlabel{fig:spin_glass_energy_2}(a), and larger deviations in the occupation of the low-energy states in \figlabel{fig:spin_glass_energy_2}(b).

Note that the dynamics considered here---finite-time coherent driving of a closed quantum system---lies outside the framework of thermalization as described by the eigenstate thermalization hypothesis, which assumes long-time evolution under a time-independent Hamiltonian \cite{dalessio_quantum_chaos2016}. A comprehensive theoretical understanding of whether finite-time adiabatic protocols can lead to thermal statistics is, to our knowledge, an open question. Consequently, it is not straightforward to establish a general relation between the DCQS parameters and the effective temperature of the corresponding distribution, and we leave such investigations for future work. Here, we instead employ the reweighting scheme described above, which does not assume that the DCQS distributions exactly follow a Boltzmann form. 

\subsection{Classical sampling methods}
\label{subsec:classical_sampling}
The calculation of thermal expectation values can be done classically using MCMC to produce samples distributed according to $\mu(s)$ \cite{dongarra2000}. One of the most commonly used method is the Metropolis-Hastings (MH) algorithm \cite{newman1999} which, starting from a sample $s$, randomly proposes a new sample $s^\prime$ according to some proposal function $g(s^\prime|s)$, and then accepts a new sample with probability
\begin{equation}
    \min\left(1, \frac{\mu(s^\prime)}{\mu(s)} \right) = \min\left(1, e^{-\beta[E(s^\prime)-E(s)]}  \right).
    \label{eq:mh}
\end{equation}
A typical choice for $g(s^\prime|s)$, which we will adopt throughout this manuscript, is to flip a single random bit. To produce samples with more statistical independence and to increase the parallelization of the method, multiple instances of MH, known as walkers, can be run in parallel, starting from a different random initial state.

While the MH algorithm can efficiently generate Boltzmann distributed samples at intermediate and high temperatures, its autocorrelation time diverges at low temperatures, making it more and more difficult to apply. Indeed, as sketched in \figlabel{fig:sketch}(a), it has been shown that the escape time from a local minimum scales as $e^{\beta\Delta E}$ and diverges at low temperatures \cite{newman1999}. Here, $\Delta E$ is the energy barrier under the allowed dynamics, i.e., local bit flips. Furthermore, systems with a large $N$ are more likely to display large Hamming distances, thus large energy barriers between local minima. 

One of the state-of-the-art classical approaches to overcome this limitation is parallel tempering (PT) \cite{hukushima1996, earl2005}, sketched in \figlabel{fig:sketch}(b). Since the MH algorithm does not get trapped in local minima at high temperatures, PT runs $n_\text{replicas}$ individual MH simulations, denoted as replicas, in parallel, each one sampling the Boltzmann distribution at a different inverse temperature 
\begin{equation}
    \beta_1,\leq \beta_2 \leq\dots\leq \beta_{n_\text{replicas}}.
\end{equation}
After every $N$ local updates proposed by the MH algorithm [\eqlabel{eq:mh}], PT attempts to swap the states of neighboring replicas at inverse temperatures $\beta_i$ and $\beta_{i+1}$. The proposed swap is accepted with probability
\begin{equation}
    \min\left(1, \exp{\left(E(s_i) - E(s_{i+1})\right)\left(\beta_i-\beta_{i+1}\right) }  \right),
    \label{eq:pt_acceptance_criteria}
\end{equation}
where $s_i$ here denotes the current state of the $i$-th replica. 
The exchange between replicas allows a colder Markov chain to escape a local minimum by swapping its state with a hotter one. In the spirit of Refs.~\cite{katzgraber2006,vousden2016,rozada2019}, we systematically optimize the number of replicas and their temperatures using an adaptive approach based on the acceptance ratio between neighboring replicas, see \applabel{app:pt} for details.

\subsection{Scalable figure of merit for low-temperature Boltzmann sampling}
\label{subsec:logz}

In this section, we introduce a scalable figure of merit that can be efficiently computed also for large values of $N$. It allows us to (i) compare the performance of DCQS against classical methods and (ii) determine whether we have sampled a sufficient number of states to accurately compute low-temperature observables.

Let us denote by $\tilde{S}$ the set of states that have been identified either by running DCQS or a classical method such as MH or PT, and let us consider the corresponding reweighed distribution $\tilde{\mu}({s})$ defined in \eqlabel{eq:tilde_p_def}.
As a metric for the quality of the reweighed distribution $\tilde{\mu}({s})$, we consider how much it deviates from the \textit{exact} Boltzmann distribution $\mu({s})$ both in terms of the commonly employed Kullback-Leibler (KL) divergence $\mathcal{D}(q||p)$ \cite{kullback1951} and the total variation distance $\delta(q,p)$~\cite{lecam1973}, where $q$ and $p$ are two arbitrary distributions. These two metrics are defined as 
\begin{equation}
\begin{aligned}
    \mathcal{D}(q||p) &= \sum_{{s}\in \mathcal{S}} q({s}) \ln\left(\frac{q({s})}{p({s})}\right), \\
     \delta(q,p) &= \frac{1}{2} \sum_{{s}\in \mathcal{S}} |q({s}) -p({s})|.
\end{aligned}
\label{eq:kl_tvd_def}
\end{equation}
While only $\delta(q,p)$ is a proper statistical distance satisfying $0 \leq \delta(q,p) \leq 1$, $\mathcal{D}(q||p)$ is lower-bounded by zero, and both quantities are zero if and only if the two distributions are identical. This makes them useful measures of similarity between distributions.
As proven in \applabel{app:fig_merit}, the distances between the reweighed distribution $\tilde{\mu}({s})$ and the exact Boltzmann distribution $\mu({s})$ can be expressed as
\begin{equation}
\begin{aligned}
    \mathcal{D}(\tilde{\mu}||\mu) &=  \ln\mathcal{Z} - \ln\mathcal{\tilde{Z}}, \\
    \delta(\tilde{\mu},\mu) &= 1 - \frac{\tilde{\mathcal{Z}}}{\mathcal{Z}}.
\end{aligned}
\label{eq:kl_tilde_p}
\end{equation}

Crucially, we notice that, regardless of the value of the exact partition function $\mathcal{Z}$, the larger $\ln \tilde{\mathcal{Z}}$, the smaller are both the KL divergence and the total variation distance. 
Therefore, given the set of states $\tilde{\mathcal{S}}_\text{DCQS}$ identified with DCQS and the sets of states $\tilde{\mathcal{S}}_\text{MH}$ and $\tilde{\mathcal{S}}_\text{PT}$ identified, respectively, with the classical MH and PT methods, we can rigorously state which method produces a distribution closer to the exact Boltzmann distribution simply by determining which one has the largest value of $\ln\tilde{\mathcal{Z}}$. Notice that while $\mathcal{Z}$ cannot be, in general, computed for large $N$ as it requires summing $2^N$ terms, in the low-temperature regime, $\tilde{\mathcal{Z}}$ only requires summing over a polynomial number of states.
We further notice that the empirical distribution produced by MH or PT is always further away from the exact Boltzmann distribution than the corresponding reweighed distribution (see \applabel{app:fig_merit} for a proof). Therefore, to ensure a fair comparison between the methods, we consider the reweighed distribution both for DCQS and for the classical sampling methods.

Additionally, we can use $\ln \tilde{\mathcal{Z}}$ to empirically determine whether we have identified the relevant states necessary to produce the Boltzmann distribution at a given target temperature. Indeed, if the value of $\ln \tilde{\mathcal{Z}}$ keeps increasing as we collect new samples (by more DCQS iterations and/or classical sampling), it means that the reweighed distribution is still far from the exact Boltzmann distribution. Conversely, if the value of $\ln \tilde{\mathcal{Z}}$ reaches a plateau and no longer increases upon the inclusion of additional states, one can be reasonably confident that the distribution has converged to the exact Boltzmann distribution. Additionally, the convergence to the Boltzmann distribution using a polynomial number of states is an empirical way of determining which temperatures are compatible with the low-temperature regime. 
Indeed, as we show in the next section, we observe that there is a temperature below which the value of $\ln \tilde{\mathcal{Z}}$ does not change increasing the number of samples. This both signals convergence to the Boltzmann distribution and identifies the low-temperature regime.

\section{Results}
\label{sec:results}
In this section, we present results obtained from running DCQS on both classical simulators and quantum hardware, specifically on the \textsc{IBM Marrakesh} and \textsc{IBM Fez} 156-qubit processors based on the heavy-hex lattice of the \textsc{IBM Heron} architecture \cite{ibm}.

In order to obtain an algorithm that can be run on present-day NISQ devices, we limit the depth of the quantum circuits, setting $n_\text{Trot}=2$ Trotter steps. We choose a control with $\dot{\lambda}_0=\dot{\lambda}_\tau=0$, such that only the $k=1$ term in \eqlabel{eq:trott} contributes. 
As detailed in \applabel{app:cd}, we compute the CD term using the first-order expansion described in \seclabel{subsec:dcqs}, which amounts to only computing the first-order coefficient $\alpha_1(\lambda)$. 
Additionally, we consider the so-called impulse regime \cite{cadavid2025}, where $|\lambda_t| \ll |\alpha_1(t)\dot{\lambda}_t|$. In this regime, the adiabatic gauge potential term $\dot{\lambda}_t\,\hat{A}_{\lambda_t}$ in \eqlabel{eq:h_ad_cd} dominates over the adiabatic term $\hat{H}_\text{ad}(\lambda_t)$, so we neglect $\hat{H}_\text{ad}(\lambda_t)$ altogether, and we only build the quantum circuit from $\dot{\lambda}_t\,\hat{A}_{\lambda_t}$. 

In \seclabel{subsec:ising}, we illustrate and validate our method by studying a 1D Ising chain with random local fields and nearest-neighbor interactions with $N=18$ and $N=124$ sites, which can be solved exactly using the transfer matrix approach \cite{yeomans1992}.
We find a good agreement of the magnetization, connected spin-spin correlation function, and average energy in the low-temperature regime, accompanied by a converged figure of merit. We then consider a three-body Hamiltonian with $N=156$ qubits in \seclabel{subsec:3body}, showing that the distributions obtained with DCQS are substantially closer to the exact low-temperature Boltzmann distribution than MH or PT per equal number of sampled states. 
Finally, we show that PT requires three orders of magnitude more samples than DCQS executed on \textsc{IBM Fez} to achieve the same quality as the DCQS distribution, corresponding to a $\sim 2\times$ runtime advantage with respect to an optimized single-core PT implementation executed on a classical computer (see App.~\ref{app:pt} for details on the classical runtime).

\subsection{Benchmarking with the 1D Ising model}
\label{subsec:ising}
We start by illustrating and validating DCQS considering the 1D Ising model
\begin{equation}
    \hat{H}_\text{f} = \sum_{i=1}^{N} \left[ h_i \, \hat{Z}_i + J_{i} \, \hat{Z}_{i} \hat{Z}_{i+1}  \right],
    \label{eq:ising_h}
\end{equation}
with $\hat{Z}_{N+1} = \hat{Z}_1$, where $h_i$ are local fields and $J_i$ are coupling constants uniformly sampled in the $[-1,1]$ interval. This model can be solved exactly using the transfer matrix approach \cite{yeomans1992}. It reduces the calculation of the partition function, and thus of thermal observables, to the trace of the product of $N$ matrices of size $2\times 2$ and to its derivatives (see \applabel{app:1d_ising} for details).

\subsubsection{Small system size: state-vector simulations}

\begin{figure}[!tb]
    \centering
    \includegraphics[width=0.99\columnwidth]{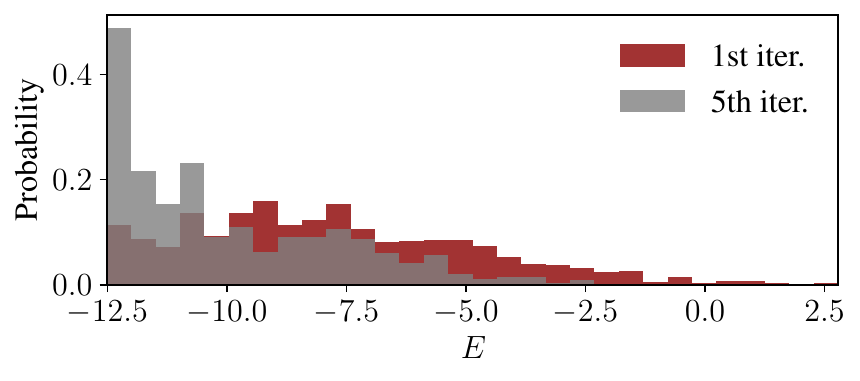}
    \caption{Energy distribution obtained from DCQS for the 1D Ising model in \eqlabel{eq:ising_h} for $N=18$ using a state-vector simulation. The red distribution corresponds to the first iteration with $b_i^{(1)}=0$ and the gray distribution to the last BF iteration with iteratively updated $b_i^{(5)}$. The parameters are $n_\text{iter} =5$, $n_\text{shots} = 1000$, $w=0.5$, and $n_\text{cvar}=20$.}
    \label{fig:ising_18_energy}
\end{figure}
First, we run DCQS using the exact state-vector simulator provided by IBM Qiskit \cite{javadiabhari2024} for $N=18$, choosing $n_\text{iter}=5$ BF iterations, setting the BF weight $w=0.5$, and collecting $n_\text{shots}=1000$ shots (samples) per iteration, so a total of $5000$ shots. 
In \figlabel{fig:ising_18_energy}, we plot the distribution of the energies found in the first iteration (red) and in the fifth iteration (gray); intermediate iterations are qualitatively similar to the fifth one. As expected, the application of the BF systematically shifts the energy distribution towards lower values (left), thereby improving the sampling of low-energy states. As described in \seclabel{subsec:dcqs}, we use all of these states $\tilde{S}_\text{DCQS}$ to compute physical observables according to \eqlabel{eq:o_tilde_avg}. 

\begin{figure}[!tb]
    \centering
    \includegraphics[width=0.99\columnwidth]{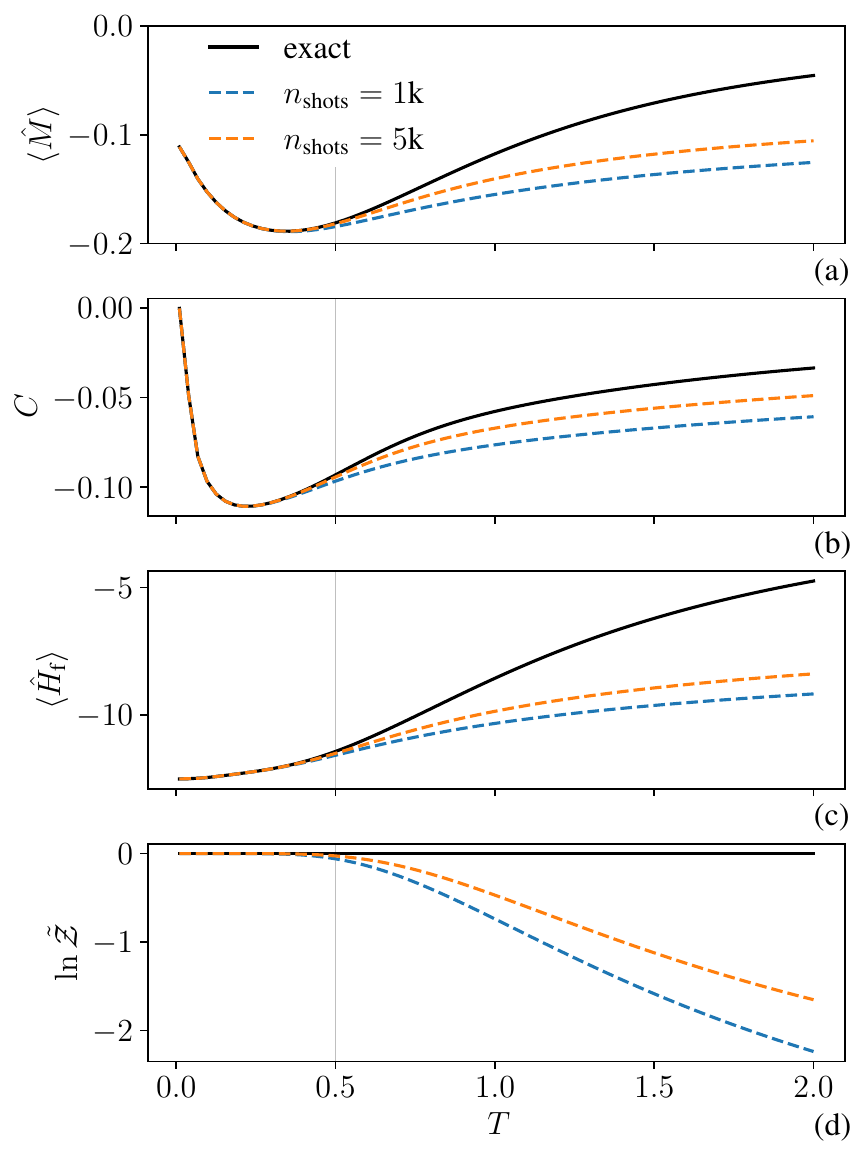}
    \caption{Observables computed for the 1D Ising model with $N=18$: (a) magnetization, (b) the connected correlation function of \eqlabel{eq:correlation_function}, (c) average energy, and (d) figure of merit $\ln \tilde{\mathcal{Z}}$ as functions of temperature. The black full lines correspond to exact results based on transfer matrix calculations. The dashed lines correspond to DCQS with the same parameters as in \figlabel{fig:ising_18_energy}, but with a number of shots per iteration reported in the legend. In panel (d) the exact partition function (black line) is used as the zero reference, while 
    the dashed lines are the corresponding values of $\ln \tilde{\mathcal{Z}}_\text{DCQS}$. The difference between the full and dashed lines corresponds to the KL diverge between the reweighed DCQS and the exact Boltzmann distribution [see \eqlabel{eq:kl_tilde_p}]. 
    The gray vertical line denotes the low-temperature regime.}
    \label{fig:ising_18_obs}
\end{figure}
In \figlabel{fig:ising_18_obs}, we benchmark our method by computing the average magnetization [panel (a)]
\begin{equation}
    \ev*{\hat{M}} = \frac{1}{N}\sum_{i=1}^{N} \ev*{\hat{Z}_i},
\end{equation}
the average connected spin-spin correlation function [panel (b)]
\begin{equation}
    C = \frac{1}{N}\sum_{i=1}^{N} \left[\ev*{\hat{Z}_i\hat{Z}_{i+1}} - 
    \ev*{\hat{Z}_i}\ev*{\hat{Z}_{i+1}}  \right],
    \label{eq:correlation_function}
\end{equation}
the average energy $\ev*{\hat{H}_\text{f}}$ [panel (c)], and the figure of merit $\ln\tilde{\mathcal{Z}}$ [panel (d)] as functions of temperature $T$. 
The black curve corresponds to exact results using the transfer matrix approach, the dashed blue curve corresponds to the DCQS results described in \figlabel{fig:ising_18_energy}, and the orange dashed line corresponds to DCQS results obtained by increasing the shots per iteration from $n_\text{shots}=1000$ to $n_\text{shots}=5000$, for a total of $25\,000$ shots.
In panel (d), the dashed lines correspond to $\ln \tilde{Z}_\text{DCQS}$ computed from the $\tilde{S}_\text{DCQS}$ states, and the black line corresponds to the exact partition function $\ln\mathcal{Z}$. Notice that the distance between the solid line and the dashed lines in \figlabel{fig:ising_18_obs}(d) is precisely the KL divergence between the DCQS reweighed distribution $\tilde{\mu}(s)$ and the exact Boltzmann distribution ${\mu}(s)$, see \eqlabel{eq:kl_tilde_p}. Since only the difference has a physical meaning, in \figlabel{fig:ising_18_obs}(d) we shift the values of the figure of merit, setting the exact $\ln\mathcal{Z}$ as the zero reference.

As we can see, DCQS perfectly matches the exact results at low temperatures, with deviations starting around $T \approx 0.5$ (gray vertical line). Notably, not only does our method capture the $T\rightarrow 0 $ limits of the observables, but it also correctly captures the non-trivial low-temperature dependence of the magnetization $\ev*{\hat{M}}$ and correlation function $C$, which display a minimum around $T\approx 0.25$.

\begin{figure}[!tb]
    \centering
    \includegraphics[width=0.99\columnwidth]{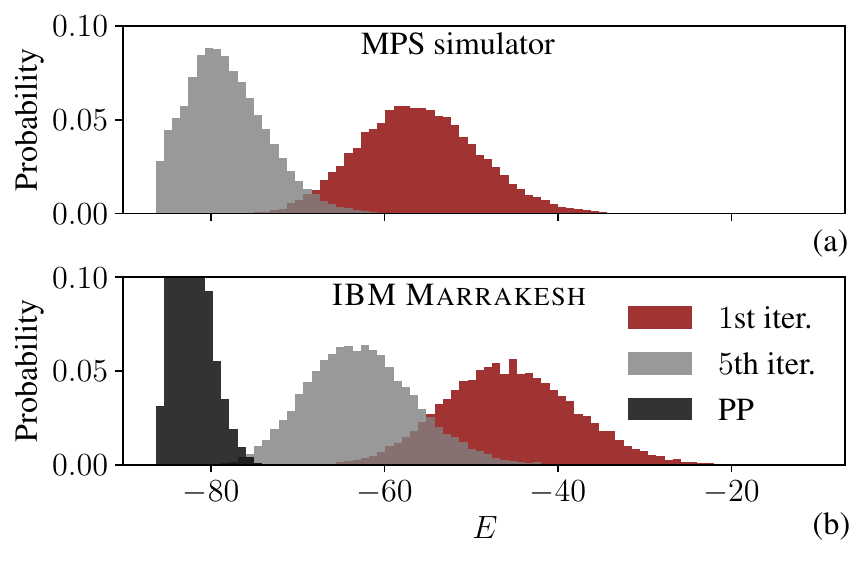}
    \caption{Energy distribution obtained running DCQS for the 1D Ising model in \eqlabel{eq:ising_h} for $N=124$ using the MIMIQ MPS simulator [panel (a)], and running on \textsc{IBM Marrakesh} [panel (b)]. The red distribution corresponds to the $1$st iteration, the gray distribution to the last BF iteration, and the black one to the states found applying post-processing to the IBM hardware data. DCQS parameters are $n_\text{iter}=5$, $n_\text{shots}=20\,000$, $w=1$, $n_\text{cvar}=20$, $n_\text{PP}=100$, and $n_\text{sweeps}=10$.}
    \label{fig:ising_124_energy}
\end{figure}
Interestingly, from \figlabel{fig:ising_18_obs}(d), we see that increasing the number of samples by a factor $5$ does not change the value of $\ln\tilde{\mathcal{Z}}_\text{DCQS}$ for $T<0.5$, whereas it increases for $T>0.5$. 
As argued in \seclabel{subsec:logz}, this is precisely the scalable criterion that we introduce to determine whether our method has converged, and to determine the low-temperature regime.
In this case, knowing the exact solution (black curves), we can indeed validate that our results have converged and produce accurate observables in the low-temperature regime.

\subsubsection{Large system size: MPS simulations and experiments}

We now push DCQS to a substantially larger regime, studying $N=124$ qubits. Given the size of the Hilbert space, we cannot perform 
exact state-vector simulations and instead, we use a matrix product state (MPS) simulator, MIMIQ, developed by QPerfect \cite{qperfect}. We also run our algorithm on \textsc{IBM Marrakesh} using an embedding of the 1D chain
into a subgraph of the device. We schedule the two-qubit Ising interactions via an edge coloring of the coupling graph, enabling two parallel layers per Trotter step that correspond to the even and odd bonds of the 1D chain. This hardware-aware schedule minimizes two-qubit gate depth on the heavy-hex architecture.

\begin{figure}[!tb]
    \centering
    \includegraphics[width=0.99\columnwidth]{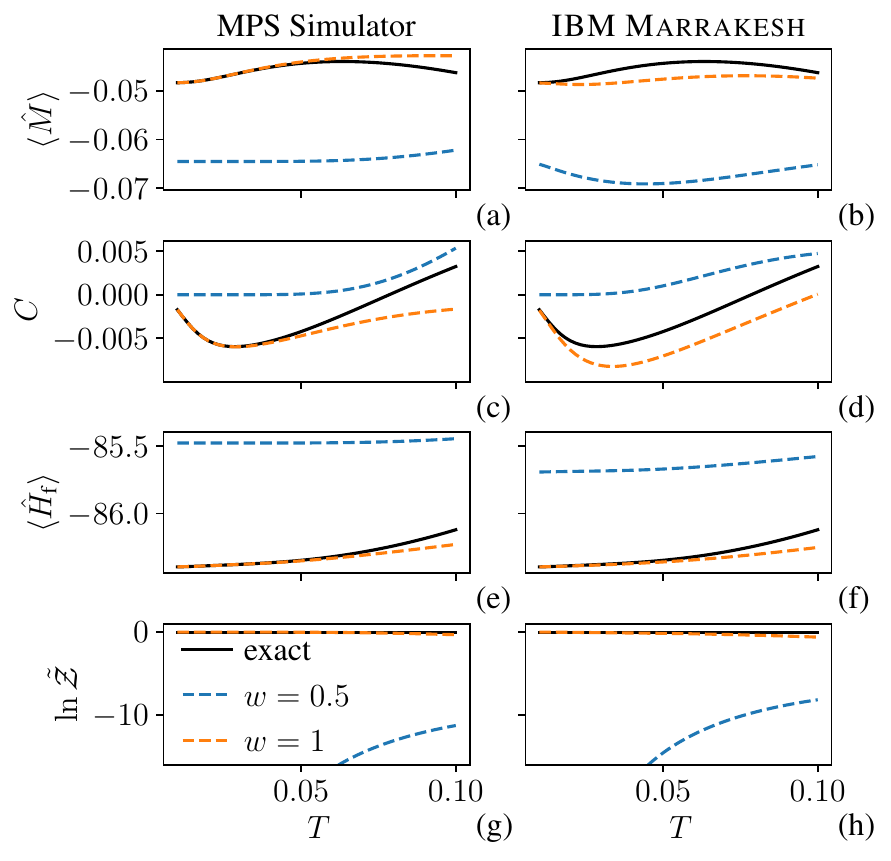}
    \caption{Observables computed for the 1D Ising model with $N=124$ and random coefficients: (a,b) magnetization, (c,d) the correlation function $C$, (e,f) average energy, and (g,h) $\ln\tilde{\mathcal{Z}}$, 
    as functions of temperature. Left panels refer to MPS simulations using the MIMIQ software~\cite{qperfect}, and the right column corresponds to data from \textsc{IBM Marrakesh} with PP. The black full lines are exact transfer matrix results. The dashed lines correspond to DCQS with the same parameters as in \figlabel{fig:ising_124_energy}, except for the BF weight $w$ reported in the legend. The $w=1.0$ curves correspond to the distributions in \figlabel{fig:ising_124_energy}.
    In panels (g,h) the exact partition function (black curve) is used as the zero reference, since only the difference is related to the KL divergence between the reweighed and exact Boltzmann distributions. 
    }
    \label{fig:ising_124_obs}
\end{figure}
In \figlabel{fig:ising_124_energy}, we plot the energy distributions, in the same style as in \figlabel{fig:ising_18_energy}, obtained by running DCQS with $n_\text{iter}=5$, $n_\text{shots}=20\,000$, and $w=1$. Both the MPS simulator [\figlabel{fig:ising_124_energy}(a)] and \textsc{IBM Marrakesh} [\figlabel{fig:ising_124_energy}(b)] results show a stark shift of the energy distribution towards lower energies with increasing BF iteration (compare the red and gray distributions), demonstrating the effectiveness of the BF at finding low energy states. 
However, due to hardware noise, we notice that the distributions from \textsc{IBM Marrakesh} are shifted towards higher energies compared to the MPS simulator, thus failing at determining the lowest energy states.

As done e.g. in Ref.~\cite{chandarana2025}, we mitigate hardware noise by applying an efficient greedy post-processing (PP) technique. Among all identified samples, we consider the $n_\text{PP}$ lowest-energy ones. For each sample, we randomly flip one bit and add the new bitstring to the set $\tilde{\mathcal{S}}_\text{DCQS}$. If the energy is lower, we update the current state; otherwise, we keep the previous one. We repeat this for a given number of sweeps $n_\text{sweeps}$, corresponding to randomly flipping $N$ bits. The black distribution shown in \figlabel{fig:ising_124_energy}(b) corresponds to the energy distribution of the states identified with PP. As we can see, this technique helps to counter hardware noise, effectively allowing us to identify the missed lowest-energy states. 
Indeed, assuming that the hardware noise leads to a few randomly flipped bits, PP is able to identify them and flip them back to a lower-energy state. For the actual implementation of PP, we apply the MH algorithm at a very low temperature, $T=0.02$, since the MH acceptance criteria in \eqlabel{eq:mh} reduces to accepting only lower energies in the zero-temperature limit.

In \figlabel{fig:ising_124_obs}, we plot the same observables as in \figlabel{fig:ising_18_obs}, with $N=124$. We compare the MPS simulator results [left column, corresponding to \figlabel{fig:ising_124_energy}(a)], and results found using \textsc{IBM Marrakesh} with PP [right column, corresponding to \figlabel{fig:ising_124_energy}(b)]. 
From \figlabel{fig:ising_124_obs}(g,h), we see that for low temperatures, the KL divergence between the reweighed distribution found with DCQS and the exact Boltzmann distribution is almost zero for $w=1$ (difference between the dashed orange and the full black curves), whereas it exhibits a large value for $w=0.5$ (difference between the dashed blue and full black curves). This indicates that, in this case, a larger BF weight $w$ is crucial to correctly capture the relevant low-energy states. This holds both for the MPS results and the quantum hardware results with PP. 
Accordingly, all observables [\figlabel{fig:ising_124_obs}(a-f)] computed with $w=1$ are in agreement with the reference results at low temperatures, while the $w=0.5$ results show discrepancies. Interestingly, DCQS with $w=1$ correctly captures the non-trivial low-temperature dependence of the observables, such as the minimum of the correlation function $C$ around $T\approx 0.04$, the increase of the average energy $\ev*{\hat{H}_\text{f}}$, and the roughly constant behavior of the magnetization $\ev*{\hat{M}}$.

Crucially, even in the absence of a reference result, we know that the results with $w=1$ are more accurate than the ones with $w=0.5$, since the $w=1$ case exhibits a substantially larger $\ln\tilde{\mathcal{Z}}$ [see \figlabel{fig:ising_124_obs}(g,h)]. 
We will use this criterion in the next section to rigorously compare the performance of DCQS against the classical MH and PT methods applied to Hamiltonians for which an analytical solution is not known.

\begin{figure}[!tb]
    \centering
    \includegraphics[width=0.99\columnwidth]{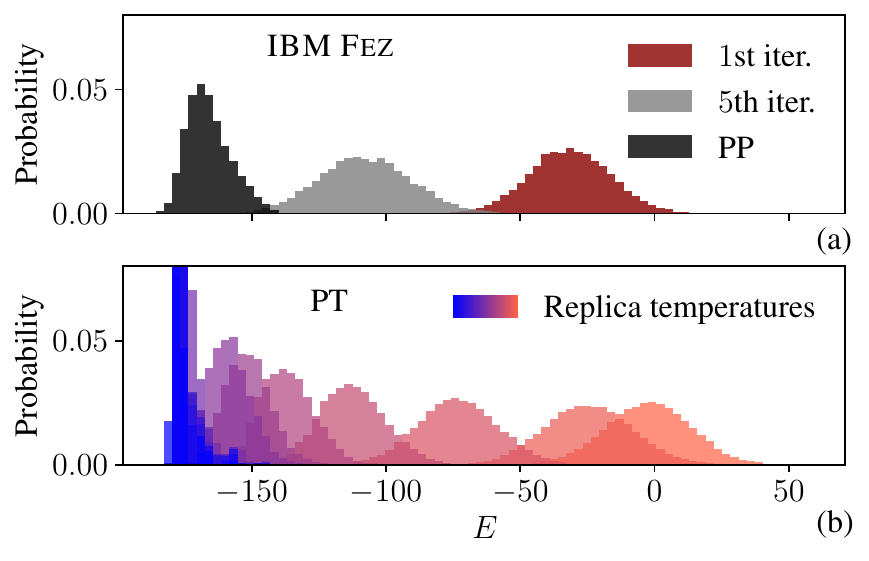}
    \caption{Energy distribution obtained for the three-body Hamiltonian of \eqlabel{eq:3body_h} with $N=156$, using (a) DCQS run on \textsc{IBM Fez}, and (b) PT. The red distribution in panel (a) corresponds to the $1$st BF iteration, the gray distribution to the last one, and the black distribution corresponds to the states found using PP. The DCQS parameters are $n_\text{iter} = 5$, $n_\text{shots} = 10\,000$, $w=10$, $n_\text{PP} = 100$, and $n_\text{sweeps}=3$. This corresponds to $50\,000$ shots generated on the QPU and $46\,800$ with PP, for a total of $96\,800$ samples. Details of the BF calculation and hardware execution are reported in \applabel{app:sidon}. (b) The same total number of samples is produced with PT using $14$ replicas with inverse temperatures chosen to produce sufficiently overlapping distributions (see \applabel{app:pt} for details). Each distribution in panel (b) corresponds to a separate replica (blue to orange is from coldest to hottest).}
    \label{fig:7}
\end{figure}

\subsection{Higher-order spin-glass Hamiltonian with $N=156$ qubits}
\label{subsec:3body}
After validating DCQS against exact results, we now apply our method to a more challenging Hamiltonian instance with $N=156$ qubits containing up to three-body terms
\begin{equation}
    \hat{H}_\text{f} = 
    \sum_{i=1}^{N}  h_i \, \hat{Z}_i + 
    \sum_{\ev*{i,j}}  J_{i,j} \, \hat{Z}_i \hat{Z}_j +
    \sum_{\ev*{i,j,k}}  K_{i,j,k} \, \hat{Z}_i \hat{Z}_j \hat{Z}_k ,
    \label{eq:3body_h}
\end{equation}
where $\ev*{i,j}$ and $\ev*{i,j,k}$ represent a summation over nearest-neighbor sites defined on the heavy-hexagonal lattice of the \textsc{IBM Heron} architecture.
In particular, we consider a Hamiltonian instance, studied in Ref.~\cite{romero_sqc2025},
with the $h_i$, $J_{i,j}$, and $K_{i,j,k}$ coefficients randomly chosen from the Sidon set $\{ \pm 8/28, \pm 13/28, \pm 19/28, \pm1 \}$~\cite{sidon1932,katzgraber2015}.
Notice that this instance contains $156$ one-, $176$ two- and $244$ three-body terms.

To evaluate the performance of DCQS with respect to classical sampling techniques, we compare it against the commonly employed MH method and the state-of-the-art low-temperature classical sampling method PT. 
The comparison is based on the figure of merit $\ln\tilde{\mathcal{Z}}$; as discussed in \seclabel{subsec:logz}, a larger value of $\ln \tilde{\mathcal{Z}}$ indicates a reweighed distribution closer to the exact Boltzmann distribution.
For an initial assessment, we generate an equal number of samples using all methods. 
Then, to truly evaluate the advantage of DCQS over classical methods for low-temperature sampling, in \seclabel{subsubsec:dcqs_adv}, we collect three orders of magnitude more samples with PT than with DCQS to identify the break-even point where the figures of merit coincide. Furthermore, we compare the runtimes of the two methods.

\subsubsection{Comparison with an equal number of samples}

We first examine the case where all methods produce an equal number of samples, including the states produced by applying PP to the IBM hardware outputs. In \figlabel{fig:7}(a), we show the energy distribution obtained from DCQS on \textsc{IBM Fez} with $n_\text{iter} = 5$ and $n_\text{shots} = 10\,000$, corresponding to a total of $50\,000$ samples generated by the QPU. The black distribution is obtained with PP with only $n_\text{PP} = 100$ and $n_\text{sweeps} =3$; this corresponds to producing (classically) an additional $46\,800$ states, for a total of $96\,800$ samples. We generate the same number of samples with MH and PT. 
\figlabel{fig:7}(b) displays the energy distribution of each of the $14$ PT replicas. The replica temperatures are chosen to guarantee sufficient overlap between neighboring distributions, and thus a sufficiently high acceptance ratio for state swaps between replicas (see \applabel{app:pt} for details). 

\begin{figure}[!tb]
    \centering
    \includegraphics[width=0.99\columnwidth]{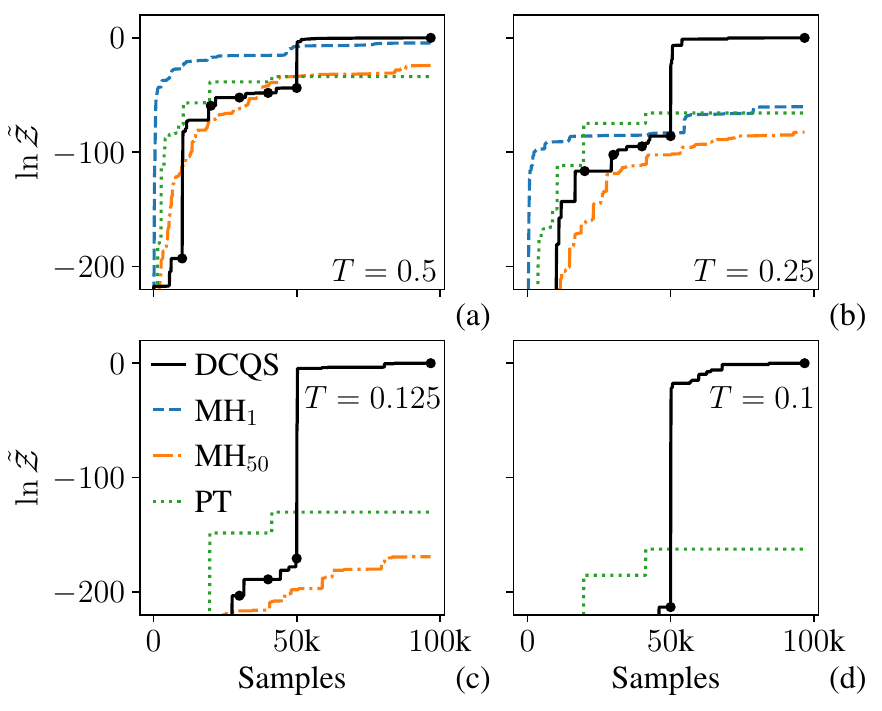}
    \caption{The cumulative figure of merit $\ln \tilde{\mathcal{Z}}$ as a function of the number of samples for the Hamiltonian of \eqlabel{eq:3body_h} with $N=156$. Here, $\ln \tilde{\mathcal{Z}}$ is computed by taking the 
    number of samples reported on the $x$-axis.  The different panels correspond to different temperatures. The black continuous line corresponds to DCQS, with the markers up to $50\,000$ delimiting the first five BF iterations. The last sector (after $50$k samples) is obtained by applying PP. The blue dashed line and the orange dashed-dotted line correspond to MH, run at the corresponding temperature, with $1$ or $50$ parallel walkers, respectively. The green dotted line corresponds to PT. A different vertical offset is applied in each panel for a clearer comparison 
    (only the difference between the curves has a physical meaning). The curves that are not visible are below the plotted range. The parameters for DCQS and PP are the ones reported in \figlabel{fig:7}. 
    }
    \label{fig:8}
\end{figure}
In \figlabel{fig:8}, we plot the \textit{cumulative} $\ln \tilde{\mathcal{Z}}$ as a function of the number of collected samples, i.e., we plot the value $\ln \tilde{\mathcal{Z}}$ would have if sampling was interrupted after collecting the number of states reported on the $x$-axis. The black continuous line corresponds to DCQS, with the first black dots up to $50\,000$ delimiting the $5$ BF iterations, and the last sector from sample $50\,000$ onward representing the states found with PP. 
The blue dashed and orange dash-dotted lines correspond to the MH method executed with $1$ ($\text{MH}_1$) and $50$ ($\text{MH}_{50}$) independent walkers. The green dotted line corresponds to PT. Notice that while MH is run separately for each temperature, both DCQS and PT are run only once to obtain the data for all temperatures, since DCQS targets low energy states directly, and PT  explores multiple temperatures in parallel, which cover the plotted range.

Notably, DCQS outperforms the classical methods at the displayed temperatures. Indeed, in all panels of \figlabel{fig:8}, the final value of $\ln \tilde{\mathcal{Z}}$ given by DCQS is the largest one, with the difference increasing as the temperature decreases. This is precisely the temperature regime targeted by DCQS, where classical methods display difficulties.

At $T=0.5$, shown in \figlabel{fig:8}(a), $\text{MH}_1$ performs best among the classical methods. 
This can be understood as follows. Since the total number of generated samples is fixed, a single walker in $\text{MH}_1$ executes more MH steps than either the $14$ replicas in PT or the $50$ walkers in $\text{MH}_{50}$.
At this relatively higher temperature, single bit flips efficiently explore the energy landscape without getting trapped in local minima, and $\text{MH}_1$ samples the relevant energy region effectively. However, as we decrease the temperature, the picture changes drastically: at $T=0.25$, in \figlabel{fig:8}(b), we see that PT performs similarly to $\text{MH}_1$. At even lower temperatures $T=0.125$ and $T=0.1$, shown in Figs. \ref{fig:8}(c) and \ref{fig:8}(d), respectively, $\text{MH}_1$ fails to sample the relevant states, and its figure of merit falls outside the plotted range. In this regime, $\text{MH}_\text{50}$ surpasses $\text{MH}_\text{1}$, yet PT emerges as the best-performing classical method by a margin. This behavior is a clear signature of the diverging escape time characteristic of MH sampling at low temperatures, discussed in \seclabel{subsec:classical_sampling}. At such low temperatures, MH walkers are increasingly likely to get stuck in local minima, and the single walker of $\text{MH}_1$ has a high probability of getting trapped. To some extent, this can be mitigated by a larger number of walkers. However, they do not restore equilibration between local minima: at $T=0.125$, $\text{MH}_\text{50}$ is still visible, but it disappears from the plot at the lowest temperature $T=0.1$. 
In contrast, PT is designed to mitigate the diverging escape time of MH, and indeed performs substantially better at low temperatures. 

\begin{figure}[!tb]
    \centering
    \includegraphics[width=0.99\columnwidth]{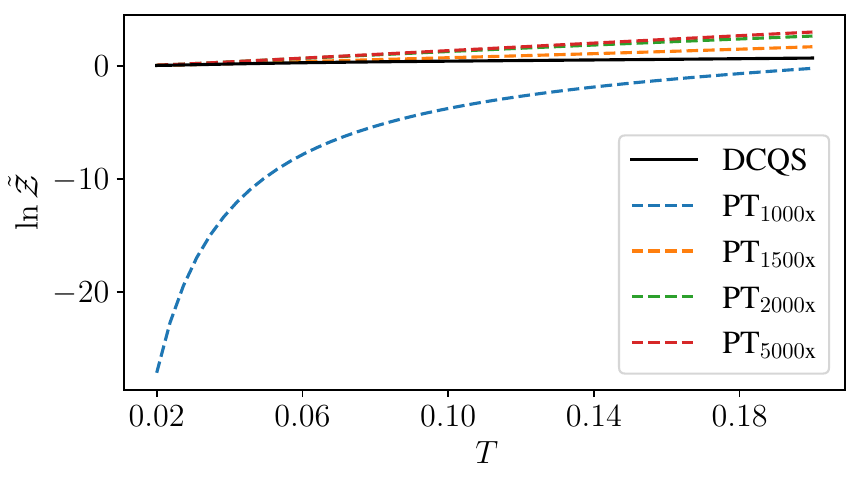}
    \caption{
    The figure of merit $\ln \tilde{\mathcal{Z}}$ for the three-body Hamiltonian in \eqlabel{eq:3body_h} with $N=156$, as a function of temperature. The black continuous line corresponds to the DCQS results shown in Figs.~\ref{fig:7} and~\ref{fig:8}. 
    The colored dashed lines correspond to PT. Each is evaluated after producing a number of states corresponding to the value reported in the legend, multiplied by the samples produced with the QPU running DCQS, i.e. $50\,000$. The DCQS parameters are reported in the caption of \figlabel{fig:7}; PT was run using the number of replicas and corresponding temperatures that were determined through the adaptive approach described in \seclabel{subsec:classical_sampling}, and detailed in \applabel{app:pt}. The acceptance ratio was set to $r=0.4$ since this value yielded the largest $\ln\tilde{\mathcal{Z}}$ at low temperatures.
    }
    \label{fig:9}
\end{figure}
Figure~\ref{fig:8} further illustrates the impact of introducing a BF and low-overhead PP within DCQS. In Figs.~\ref{fig:8}(a,b), we see an abrupt increase of $\ln \tilde{\mathcal{Z}}$ immediately after the first marker at $10\,000$ samples, marking the onset of the first iteration with a non-null BF. Furthermore, in all panels of \figlabel{fig:8}, an even larger increase of $\ln \tilde{\mathcal{Z}}$ occurs after the $50\,000$-sample point, corresponding to the effect of PP. 

\subsubsection{Sampling and runtime advantage}
\label{subsubsec:dcqs_adv}
We now want to rigorously assess the advantage of DCQS over classical methods in the low-temperature regime in terms of the number of samples and runtime. 
To this end, we choose the best classical method, PT, and we use it to produce three orders of magnitude more samples than the quantum samples ($50\,000$) produced by DCQS. Our aim is to identify when PT can match the figure of merit $\ln \tilde{\mathcal{Z}}$ obtained with DCQS.

\begin{figure}[!tb]
    \centering
\includegraphics[width=0.99\columnwidth]{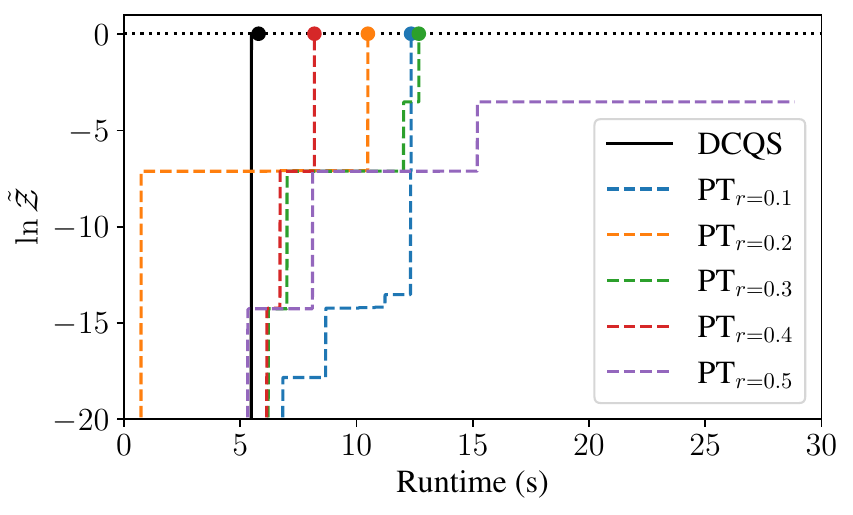}
    \caption{Cumulative figure of merit $\ln \tilde{\mathcal{Z}}$ as a function of runtime for $T=0.02$. The black continuous line corresponds to DCQS, and the colored dashed lines to PT with the acceptance ratio $r$ reported in the legend. The horizontal dotted black line marks the final value of $\ln \tilde{\mathcal{Z}}$ achieved by DCQS (black circle), and the colored circles indicate when PT matches the $\ln \tilde{\mathcal{Z}}$ of DCQS. The parameters for DCQS are reported in the caption of \figlabel{fig:7}, and the parameters for PT, described in \seclabel{subsubsec:dcqs_adv}, are detailed in \applabel{app:pt}.}
    \label{fig:10}
\end{figure}

To rigorously compare DCQS and PT, we proceed as follows. 
Since PP is stochastic in nature, we repeat the PP of the samples obtained from \textsc{IBM Fez} 50 times, choosing all integers in the $[0,49]$ interval as seeds for the random number generator. For each seed, we apply two PP sweeps to the $2000$ lowest-energy states. 
As detailed in \applabel{app:sidon}, we find large values of $\ln \tilde{\mathcal{Z}}$ at low temperatures $9$ times out of $50$, corresponding roughly to a $20\%$ probability. In Figs.~\ref{fig:9} and~\ref{fig:10}, we display the results obtained with one of these $9$ PP initializations; however, all lead to the same qualitative behavior and conclusions.
To compare with PT, we employ an adaptive scheme to systematically determine the optimal number of replicas and their temperatures, based on reaching a target acceptance ratio $r$ 
between neighboring replicas (see \seclabel{subsec:classical_sampling} and \applabel{app:pt} for details). We repeat the PT optimization for $5$ different values of the acceptance ratio $r=\{0.1, 0.2, 0.3, 0.4, 0.5\}$, finding that $r=0.4$ yields the largest figure of merit at low temperatures.

In \figlabel{fig:9}, we plot $\ln \tilde{\mathcal{Z}}$ as a function of temperature, computed using DCQS (black full line), and PT (dashed colored lines). The PT is obtained with the best-performing acceptance ratio $r=0.4$. Each color corresponds to a different number of samples produced with PT, with the number reported in the legend indicating how many more samples were produced compared to DCQS's quantum samples ($50\,000$). For example, $\text{PT}_{1000\text{x}}$, the blue dashed line, corresponds to producing $50\,000 \times 1000 = 50$ million samples with PT. 

Notably, even after the hyperparameter optimization, PT still requires between $1000$ (blue curve) and $1500$ (orange curve) times more samples than DCQS to achieve a comparable figure of merit, corresponding to three orders of magnitude more samples. This observation highlights an algorithmic advantage of DCQS over classical sampling approaches. 
From \figlabel{fig:9}, we further confirm that DCQS reaches convergence at $T<0.06$, since the curves of DCQS, $\text{PT}_{2000\text{x}}$, and $\text{PT}_{5000\text{x}}$ all coincide. More specifically, their values are almost identical in this low-temperature regime, while for $T>0.06$, they start to deviate.
These deviations are attributable to the gradual contribution of more and more excited states to the Boltzmann distribution at higher temperatures.
The overlap at $T<0.06$ suggests that the system is indeed in the low-temperature regime, where the Boltzmann distribution can be accurately reproduced using only a fraction of the states in the $2^{156} \approx 10^{47}$-dimensional Hilbert space.

In \figlabel{fig:10}, we plot the cumulative $\ln \tilde{\mathcal{Z}}$ reached by DCQS (black thick line), and by PT (dashed colored lines) at $T=0.02$ in the low-temperature regime, as functions of the runtime of the corresponding algorithm. Each dashed line corresponds to a different value of the acceptance ratio $r$ indicated in the legend. The horizontal dotted black line marks the final value of $\ln \tilde{\mathcal{Z}}$ achieved by DCQS, and the colored circles mark the runtimes at which PT reaches the same value.
The runtime of PT is evaluated using an optimized C++ implementation running on a single core, which can generate $8.7$ million samples per second (see App.~\ref{app:pt} for details). 
For DCQS, it is possible to get $10\,000$ shots per second from IBM quantum hardware by setting a suitable repetition delay~\cite{javadiabhari2024}. Then, PP can generate $8.7$ million samples per second as PT. 
Since PP leads to a high value of $\ln \tilde{\mathcal{Z}}$ with a $20\%$ probability, we multiply the PP time by a factor $5$, and we account for the runtime necessary for the calculation of the BF (see \applabel{app:sidon} for details).
We notice, however, that the time for PP and for the BF calculation is negligible compared to the $5$ seconds necessary to generate the $50\,000$ quantum samples.

Remarkably, all five independent runs of PT require more time than DCQS to reach the same figure of merit, indicating a runtime advantage of the DCQS approach. Indeed, while DCQS takes approximately $5$ seconds to reach the final value of $\ln \tilde{\mathcal{Z}}$, $\text{PT}_{r=0.4}$ takes around $8$ seconds, $\text{PT}_{r=0.2}$ $10$ seconds, $\text{PT}_{r=0.1}$ and $\text{PT}_{r=0.3}$ $12$ seconds, and $\text{PT}_{r=0.5}$ does not reach the target accuracy after $30$ seconds. The runtime of DCQS is therefore smaller by approximately a factor $2$. As detailed in \applabel{app:sidon}, we find that DCQS has a runtime quantum advantage over all $5$ runs of PT also at $T=0.1$, while $\text{PT}_{r=0.2}$ slightly surpasses the runtime of DCQS at $T=0.2$. Indeed, it is expected that, as the temperature increases, classical methods become more efficient and thus faster.

\section{Conclusions}
\label{sec:conclusions}
In this work, we introduce DCQS, a quantum-classical algorithm to efficiently estimate the low-temperature Boltzmann distribution of arbitrary classical Hamiltonians. Based on a combination of digitized counterdiabatic techniques, adaptive bias fields, and classical reweighting, DCQS is tailored to suitably work on current NISQ devices, where noise and limited coherence times are critical constraints. 

To quantify convergence and performance, we define a scalable figure of merit and benchmark DCQS
against classical samplers, including the Metropolis-Hastings (MH) algorithm and parallel tempering (PT), the state-of-the-art method for low-temperature sampling.
We first validate our approach by computing low-temperature observables of the disordered one-dimensional Ising model with nearest-neighbor interactions, 
for systems of $18$ and $124$ qubits. By performing state-vector and
MPS simulations
and experiments on the \textsc{IBM Marrakesh} quantum processor, 
we reproduce the nontrivial temperature dependence of the magnetization, spin-spin correlation function, and energy, in agreement with exact transfer-matrix results.

We then tackle a more challenging Hamiltonian with $156$~qubits and hundreds of one-, two-, and three-body terms. Using a hardware-aware embedding of the quantum circuit, we show that DCQS outperforms both the MH algorithm and PT if the number of samples is fixed. That is, DCQS produces a distribution closer to the exact Boltzmann distribution as measured by the KL divergence and total variation distance. 

Finally, we rigorously assess the performance of DCQS against classical methods in the low-temperature regime using an optimized implementation of PT. We find that PT requires three orders of magnitude more samples to match the accuracy of DCQS, corresponding to a $\sim 2\times$ runtime quantum advantage on the quantum processor. Considering the remarkable success of DCQS on \textsc{IBM Fez}, a greater enhancement is expected with future hardware generations, as qubit number, quality, and connectivity continue to improve. Furthermore, our work provides a large-scale demonstration (156 qubits) of Boltzmann sampling on a digital quantum computer.

This approach opens many venues for further research. The method can be useful in characterizing low-temperature phases and phase transitions of classical models that are challenging for numerical methods. Difficulties in simulating thermal equilibrium at low temperatures may arise from disorder and frustration, such as in spin glasses~\cite{dahlberg2025}, local constraints and topological order~\cite{castelnovo2012}, anisotropy, or weak interlayer couplings~\cite{moodie2020}.

Beyond benchmarking DCQS against PT, one can envision a synergy between the two algorithms that combines their complementary strengths. For example, DCQS may be used as a global sampler, accelerating the discovery of distant local minima (in Hamming distance), while PT or other MCMC-based algorithms efficiently sample around the identified minima. Such a hybrid approach could extend DCQS to higher-temperature regimes. 
We observe empirically that the energy distributions produced by DCQS for the models studied here, in the absence of reweighting, approximate Boltzmann statistics at an effective temperature. This observation suggests a connection between DCQS and effective thermal behavior, the deeper understanding of which could enable direct Boltzmann sampling on 
digital quantum computers. 

\section{Acknowledgments}
We acknowledge the use of IBM Quantum services for this work. The views expressed are those of the authors and do not reflect the official policy or position of IBM or the IBM Quantum team.

\appendix

\section{First-order counterdiabatic driving}
\label{app:cd}
In this appendix, we describe the first-order counterdiabatic approximation used in this paper. As mentioned in the main text, and further detailed in Ref.~\cite{cadavid2025}, we expand the adiabatic gauge potential $\hat{A}_{\lambda}$ as the sum \cite{claeys2019}
\begin{equation}
    \hat{A}_\lambda^{(l)} = i\sum_{k=1}^l\alpha_k(\lambda)\,\hat{\mathcal{O}}_{2k-1}(\lambda),
    \label{eq:a_ansatz}
\end{equation}
where $l$ represents the expansion order and 
\begin{equation}
    \hat{\mathcal{O}}_k(\lambda) = [\hat{H}_\text{ad}(\lambda), \hat{\mathcal{O}}_{k-1}(\lambda)]
\end{equation} 
with 
\begin{equation}
    \hat{\mathcal{O}}_0(\lambda) = \partial_\lambda \hat{H}_\text{ad}(\lambda).
\end{equation}
To determine the choice of the $\alpha_k(\lambda)$ coefficients, we use the variational approach proposed in Ref.~\cite{sels2017}, which consists of minimizing the functional
\begin{equation}
    S(\hat{A}_\lambda) = \Tr[\hat{G}^2_\lambda(\hat{A}_\lambda)],
\end{equation}
where
\begin{equation}
    \hat{G}_\lambda(\hat{A}_\lambda) = \partial_\lambda \hat{H}_\text{ad}(\lambda) + i [\hat{A}_\lambda, \hat{H}_\text{ad}(\lambda)],
\end{equation}
considering \eqlabel{eq:a_ansatz} as an Ansatz. Since \eqlabel{eq:a_ansatz} is linear in the coefficients, and $S(\hat{A}_\lambda)$ is a quadratic functional, the minimization can be carried out analytically for each value of $\lambda$, leading to a set of linear equations for the $\alpha_k(\lambda)$ coefficients.

In order to restrict the number of gates, such that the corresponding quantum circuit can be executed on NISQ computers within their coherence time, in this manuscript, we only consider the first-order term $\hat{A}^{(1)}_\lambda$. This requires determining the single coefficient $\alpha_1(\lambda)$. Following the minimization procedure outlined above, it is given by \cite{cadavid2025}
\begin{equation}
    \alpha_1(\lambda) = - \frac{\Tr[\hat{\mathcal{O}}^\dagger_1(\lambda) \hat{\mathcal{O}}_1(\lambda)]} {\Tr[\hat{\mathcal{O}}^\dagger_2(\lambda) \hat{\mathcal{O}}_2(\lambda)]}.
\end{equation}

\section{Fitting an effective temperature to the empirical distribution}
\label{app:fit_temp}
In this section, we introduce an effective temperature $T_\text{eff}$ associated with an empirical distribution $\bar{\mu}(s)$. We define $T_\text{eff}$ as the temperature 
that minimizes the KL divergence between $\bar{\mu}(s)$ and the exact Boltzmann distribution $\mu(s)$. We then show that this definition is equivalent to determining the effective temperature $T_\text{eff}$ such that the mean energy of $\mu(s)$ matches that of $\bar{\mu}(s)$. For convenience, in this appendix, we consider the effective inverse temperature $\beta_\text{eff}$ instead of the temperature $T_\text{eff}$.

Using \eqlabel{eq:kl_bar_p}, we have that
\begin{equation}
    \mathcal{D}(\bar{\mu}||\mu) =  \ln\mathcal{Z} + \beta \ev*{E}_{\bar{\mu}} - S(\bar{\mu}).
\end{equation}
To determine the value of $\beta$ that minimizes this distance, we take its derivative in $\beta$ and set it equal to zero, leading to the condition
\begin{equation}
    -\frac{\partial \ln\mathcal{Z}}{\partial\beta} = \ev{E}_{\bar{\mu}}.
\end{equation}
Using the thermodynamic identity $-\partial \ln\mathcal{Z}/\partial\beta = \ev{E}_\mu$, where $\ev{E}_\mu$ is the average energy computed using the exact Boltzmann distribution $\mu({s})$, we find that we can determine the temperature by solving the equation
\begin{equation}
    \ev{E}_{\mu}(\beta) = \ev{E}_{\bar{\mu}},
    \label{eq:energy_fit}
\end{equation}
concluding the proof.

\begin{figure}[!tb]
    \centering
    \includegraphics[width=0.99\columnwidth]{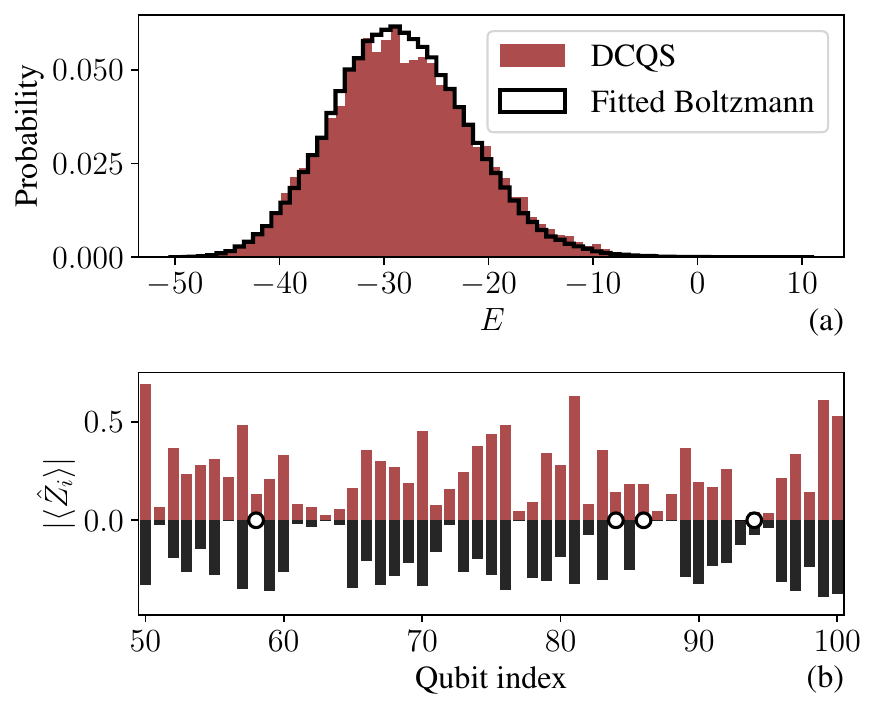}
    \caption{Comparison between the DCQS distribution for the $156$ qubit quadratic Hamiltonian $\hat{H}_\text{f}^{(1)}$ executed on \textsc{IBM Kingston} (red bars), and the Boltzmann distribution of $\hat{H}_\text{f}^{(1)}$ computed using the MH algorithm (black line and bars). The temperature of the Boltzmann distribution is fitted by matching the average energy of the DCQS and of the Boltzmann distribution. Panel (a) represents the energy distribution, and panel (b) displays $|\ev*{\hat{Z}_i}|$ for DCQS (red bars) and $-|\ev*{\hat{Z}_i}|$ for the Boltzmann distribution. The white circles denotes when the sign of the two methods does not match. DCQS is executed for a single iteration (no BF), producing $8000$ shots. The MH algorithm is executed with $10$ parallel walkers, the first $5000$ steps are removed to ensure thermalization, and then a further $45\,000$ steps per walker are produced. The fitted temperature is $T_\text{eff}=2.21$.}
    \label{fig:11}
\end{figure}

\section{DCQS as an (approximate) Boltzmann sampler}
\label{app:dcqs_boltzmann}
In this appendix, we provide further evidence that DCQS produces distributions that are similar to the exact Boltzmann distribution.
To this aim, we generate two random Hamiltonians, $\hat{H}_\text{f}^{(1)}$ and $\hat{H}_\text{f}^{(2)}$, for $156$ qubits, with local and two-body terms defined on the heavy-hexagonal lattice, and random coefficients in the $[0,1]$ interval. We then run a single iteration of DCQS (i.e., without any BF) on \textsc{IBM Kingston}, producing $8000$ shots. We build quantum circuits as described in \seclabel{sec:results} of the main text, and we do not perform any postprocessing.

In \figlabel{fig:11}, we show the results relative to $\hat{H}_\text{f}^{(1)}$. Figure~\ref{fig:11}(a) shows the energy distribution obtained from DCQS (red distribution), compared to the Boltzmann distribution of the same $\hat{H}_\text{f}^{(1)}$ Hamiltonian (black line), but computed using the MH algorithm described in the main text. In particular, we consider $10$ parallel walkers, remove the first $5000$ states produced by all walkers, to make sure that they have thermalized, and we produce a further $45\,000$ states from all walkers. The temperature of the Boltzmann distribution is fitted by matching the average energy of the DCQS distribution with the average energy of the Boltzmann distribution (see \applabel{app:fit_temp} for an explanation of this approach). Since the temperature depends monotonically on the average energy, we can efficiently perform this fit using a bisection-based approach. The value of the fitted temperature is $T_\text{eff}=2.21$.

As we can see, the energy distributions are remarkably similar. 
For further verification, in \figlabel{fig:11}(b), we compare the expectation value $\ev*{\hat{Z}_i}$ of $50$ qubits. In particular, the red bars represents the value of $|\ev*{\hat{Z}_i}|$ predicted by DCQS, whereas the black bars represent $-\ev*{\hat{Z}_i}$ predicted by sampling the Boltzmann distribution. A white circle is added when the signs predicted by the two methods do not coincide. We notice that the distributions are qualitatively quite similar, though there are some deviations both in terms of magnitude and sign.

\begin{figure}[!tb]
    \centering
    \includegraphics[width=0.99\columnwidth]{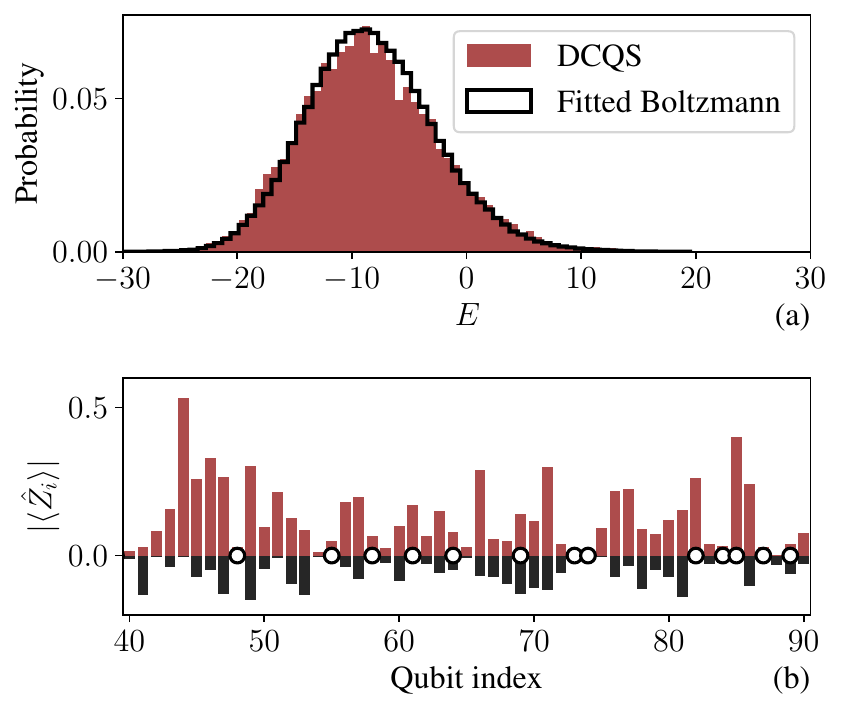}
    \caption{Same quantities as in \figlabel{fig:11}, but relative to $\hat{H}^{(2)}_\text{f}$. All other parameters are the same. The fitted temperature is $T_\text{eff}=4.76$.}
    \label{fig:12}
\end{figure}
In \figlabel{fig:12}, we plot the same quantities as in \figlabel{fig:11}, but relative to $\hat{H}^{(2)}_\text{f}$. In this case, the fitted temperature is $T_\text{eff}=4.76$.
The qualitative match between the energy distributions, and the corresponding local expectation values, are confirmed also in this case. It is interesting to observe how the fitted temperature is higher in this case than in the previous one. As a future research direction, it would be interesting to further investigate the relation between the DCQS parameters, type of quantum hardware, and fitted temperature.

\section{Parallel tempering}
\label{app:pt}
In this appendix we provide additional details on the parallel tempering implementation. In particular, in \seclabel{app:pt_replicas} we describe the adaptive procedure used to select the number of replicas and their temperatures systematically, in \seclabel{app:pt_subsec} we report the parameters used for the results reported in \seclabel{subsec:3body} of the main text, in \seclabel{app:pt_runtime} we detail the classical hardware used for the determination of classical runtime speed.

\subsection{Adaptive replica selection}
\label{app:pt_replicas}

A common criterion used in the literature is to adapt the number and temperatures of the replicas in order to guarantee a roughly uniform and sufficient acceptance ratio for exchanges between neighboring replicas ~\cite{katzgraber2006,vousden2016,rozada2019}. Indeed, if the acceptance ratio is too high, we are most likely using too many replicas, and thus wasting computational resources. If, conversely, it is too low, the mixing time will be slow, leading to slower convergence. 

By inspecting the acceptance probability in \eqlabel{eq:pt_acceptance_criteria}, we see that closely spaced inverse temperatures (i.e., small values of $|\beta_{j+1}-\beta_j|$) increase the acceptance ratio, while distant inverse temperatures (i.e., large values of $|\beta_{j+1}-\beta_j|$) lead to a lower acceptance probability. We therefore adopt an adaptive approach, inspired by Refs.~\cite{katzgraber2006,vousden2016,rozada2019}, to systematically determine the number of replicas and their inverse temperatures.

The adaptive method relies on fixing three quantities: $\beta_{\text{max}}$, corresponding to the lowest temperature we want to sample; $\beta_{\text{min}}$, corresponding to a sufficiently high temperature, such that it is easy to sample using MH with local flips; and $r$, the target acceptance ratio. 
We then perform the following iterative procedure. In the first iteration ($i=1$), we consider two replicas with fixed inverse temperatures $\{ \beta_{\text{min}}, \beta_{\text{max}} \}$.

Let $\{\beta_1,\leq \beta_2 \leq\dots\leq \beta_{n_\text{i}}\}$ be the inverse temperatures during the $i$th iteration. We run PT with these replica inverse temperatures for a given number of steps $n_\text{pt-steps}$. We then compute $r_j$ for $j=1,2,\dots, n_\text{i}-1$, corresponding the acceptance ratio between the $j$th and the $(j+1)$th replica. If all $r_j> r$, the adaptive algorithm has reached convergence, and $\{\beta_1,\leq \beta_2 \leq\dots\leq \beta_{n_\text{i}}\}$ correspond to the final identified inverse temperatures. 
Otherwise, for each $r_j<r$, we introduce a new inverse temperature $(\beta_j + \beta_{j+1})/2$ which, as argued above, is expected to increase the acceptance ratio at the next iteration. The procedure is then repeated until convergence.

\subsection{Parallel tempering parameters chosen in \seclabel{subsec:3body} of the main text}
\label{app:pt_subsec}
We now provide the details of the parallel tempering simulations performed in \seclabel{subsec:3body} of the main text. 
In all results reported in the manuscript, we fix $\beta_\text{min}=0.01$, and $\beta_\text{max}=50$. The value $\beta_\text{min}=0.01$ corresponds to a high temperature where we verify that the MH algorithm converges quickly.
The value $\beta_\text{max}=50$ corresponds to the lowest temperature reported in Figs.~\ref{fig:9} and ~\ref{fig:10} of the main text, i.e. $T=0.02$. As argued in the main text, this value is indeed in the low-temperature regime where we expect an advantage using DCQS.

In the initial assessment, reported in Figs.~\ref{fig:7} and \ref{fig:8} of the main text, we choose the intermediate temperatures manually, such that the energy distributions of neighboring replicas display a sufficient overlap. Indeed, a sufficient overlap between replicas guarantees that the acceptance ratio is not too low. The overlap can be visually seen directly in \figlabel{fig:7}(b) of the main text. The values of the inverse temperatures used in Figs.~\ref{fig:7} and~\ref{fig:8} are $\{0.01$, $0.1$, $0.3$, $0.5$, $0.7$, $0.9$, $1.0$, $3.0$, $5.0$, $7.0$, $10.0$, $20.0$, $35.0$, $50.0\}$.

For the analysis carried out in \seclabel{subsubsec:dcqs_adv}, and reported in Figs.~\ref{fig:9} and~\ref{fig:10}, we employ the adaptive method described above, repeating the optimization for $r=\{0.1, 0.2, 0.3, 0.4, 0.5\}$. For each acceptance ratio, we fix $\beta_\text{min}=0.01$, and $\beta_\text{max}=50$, and we consider $n_\text{pt-steps}= 156\,000$ to ensure sufficient statistics. All runs are initialized with the same seed $0$ for random number generation. 
The final inverse temperatures are reported in Table~\ref{tab:betas}.

\begin{table}[ht]
\centering
\small
% ↓ shrink vertical row spacing just for this tabular
{\renewcommand{\arraystretch}{1}%
\begin{tabular}{ll}
\hline
$r$ & Replica inverse temperatures \\
\hline
0.1 & 0.01, 0.11, 0.21, 0.30, 0.40, 0.50, 0.60, 0.79, 0.99, 1.18, 1.57,\\
& 3.13, 6.26, 9.38, 12.51, 18.76, 25.00, 28.13, 31.25, 37.50, 40.63,\\
& 43.75, 46.88, 50.00 \\
\hline
0.2 & 0.01, 0.11, 0.21, 0.30, 0.40, 0.50, 0.60, 0.69, 0.79, 0.99, 1.18,\\
& 1.57, 2.35, 3.13, 6.26, 9.38, 12.51, 18.76, 25.00, 28.13, 31.25,\\
& 34.38, 37.50, 40.63, 43.75, 45.31, 46.88, 48.44, 50.00 \\
\hline
0.3 & 0.01, 0.06, 0.11, 0.16, 0.21, 0.25, 0.30, 0.35, 0.40, 0.50, 0.60,\\
& 0.69, 0.79, 0.89, 0.99, 1.18, 1.38, 1.57, 1.96, 2.35, 3.13, 4.70,\\
& 6.26, 9.38, 12.51, 15.63, 18.76, 25.00, 28.13, 31.25, 34.38,\\
& 37.50, 40.63, 43.75, 46.88, 48.44, 49.22, 50.00 \\
\hline
0.4 & 0.01, 0.06, 0.11, 0.16, 0.21, 0.25, 0.30, 0.35, 0.40, 0.45, 0.50,\\
& 0.60, 0.69, 0.79, 0.89, 0.99, 1.18, 1.38, 1.57, 1.96, 2.35, 3.13,\\
& 4.70, 6.26, 9.38, 12.51, 15.63, 18.76, 21.88, 25.00, 28.13,\\
& 31.25, 34.38, 37.50, 39.06, 40.63, 43.75, 45.31, 46.88, 47.66,\\
& 48.44, 49.22, 49.61, 50.00 \\
\hline
0.5 & 0.01, 0.06, 0.11, 0.16, 0.21, 0.25, 0.30, 0.35, 0.40, 0.45, 0.50,\\
& 0.55, 0.60, 0.64, 0.69, 0.74, 0.79, 0.89, 0.99, 1.08, 1.18, 1.38,\\
& 1.57, 1.96, 2.35, 3.13, 4.70, 6.26, 7.82, 9.38, 12.51, 15.63,\\
& 18.76, 21.88, 25.00, 28.13, 31.25, 32.82, 34.38, 35.94, 37.50,\\
& 39.06, 40.63, 42.19, 43.75, 45.31, 46.88, 47.66, 48.44, 49.22,\\
& 50.00 \\
\hline
\end{tabular}%
}
\caption{Target acceptance ratio $r$ and corresponding inverse temperatures $\beta_i$.}
\label{tab:betas}
\normalsize
\end{table}

\subsection{Classical hardware runtime}
\label{app:pt_runtime}
All PT benchmarks were executed on a MacBook Pro equipped with an Apple M4 Pro processor (12-core CPU, 24 GB unified memory) using our C++/OpenMP implementation. After accounting for parallel execution with twelve threads, the single-core equivalent throughput corresponds to approximately 8.7 million spin-update attempts per second. This per-core value is used for runtime comparisons.

\section{Properties of the figure of merit}
\label{app:fig_merit}
In this appendix, we prove \eqlabel{eq:kl_tilde_p} of the main text, and we provide additional details and properties on the figure of merit.
Using the definition in \eqlabel{eq:kl_tvd_def} of the main text, we can express the KL divergence between the reweighed distribution $\tilde{\mu}({s})$ and the exact Boltzmann distribution $\mu({s})$, respectively defined in Eqs.~\eqref{eq:tilde_p_def} and~\eqref{eq:boltzmann_def} of the main text, as
\begin{align}
    \begin{split}
    \mathcal{D}(\tilde{\mu}||
\mu) &= \sum_{{s}\in \tilde{\mathcal{S}}} \tilde{\mu}({s}) \ln\left(\frac{\tilde{\mu}({s})}{\mu({s})}\right) \\
    &= \sum_{{s}\in \tilde{\mathcal{S}}} \tilde{\mu}({s})\left[ \ln\mathcal{Z} - \ln\mathcal{\tilde{Z}}  \right] =  \ln\mathcal{Z} - \ln\mathcal{\tilde{Z}},
    \end{split}
    \label{eq:kl_tilde_p_app}
\end{align}
where we use the fact that both distributions are proportional to $e^{-\beta E(s)}$, and that $\sum_{{s}\in \tilde{\mathcal{S}}} \tilde{\mu}({s})=1$. Similarly, using the definition in \eqlabel{eq:kl_tvd_def}, we can express the total variation distance between $\tilde{\mu}({s})$ and $\mu({s})$ as
\begin{align}
\begin{split}
    \delta(\tilde{\mu},\mu) &= \frac{1}{2} \sum_{{s}\in \mathcal{S}} |\tilde{\mu}({s}) -\mu({s})| \\
    &=\frac{1}{2} \sum_{{s}\in \tilde{\mathcal{S}}} \left(\frac{e^{-\beta E({s})}}{\tilde{\mathcal{Z}} }-\frac{e^{-\beta E({s})}}{\mathcal{Z}}\right) + \frac{1}{2} \sum_{{s}\in (\mathcal{S} - \tilde{\mathcal{S}})} \frac{e^{-\beta E({s})}}{\mathcal{Z}} \\
    &= 1 - \frac{\tilde{\mathcal{Z}}}{\mathcal{Z}},
    \end{split}
    \label{eq:dist_tilde_p_1_app}
\end{align}
where we denote with $(\mathcal{S}-\tilde{\mathcal{S}})$ the set of states present in $\mathcal{S}$, but not in $\tilde{\mathcal{S}}$. We further used the fact that $\tilde{\mu}(s) > \mu(s)$ for $s\in \tilde{\mathcal{S}}$, that $\tilde{\mu}(s)=0$ for $s\in (\mathcal{S} - \tilde{\mathcal{S}})$, and the definitions of $\mathcal{Z}$ and of $\tilde{\mathcal{Z}}$ to re-sum the exponentials. This concludes the proof of \eqlabel{eq:kl_tilde_p}.
Interestingly, we see that the two distances, which are in general not one-to-one-related, can be written one as a function of the other. Indeed, combining Eqs.~\eqref{eq:kl_tilde_p_app} and~\eqref{eq:dist_tilde_p_1_app}, we find that
\begin{equation}
    \delta(\tilde{\mu},\mu) = 1 - e^{-( \ln\mathcal{Z} - \ln\mathcal{\tilde{Z}} )} = 1- e^{-\mathcal{D}(\tilde{\mu}||\mu)}.
    \label{eq:dist_tilde_p_2}
\end{equation}

As mentioned in the main text, we now show that any empirical distribution $\bar{\mu}(s)$, which is non-null for a set of states $\tilde{\mathcal{S}}$, is further away from the exact Boltzmann distribution that the corresponding reweighed distribution $\tilde{\mu}(s)$.
Mathematically, we wish to prove that
\begin{align}
    \mathcal{D}(\tilde{\mu}||\mu) &\leq  \mathcal{D}(\bar{\mu}||\mu),
    \label{eq:emp_kl}\\
    \delta(\tilde{\mu},\mu) &\leq  \delta(\bar{\mu},\mu).
    \label{eq:emp_tvd}
\end{align} 
This implies that any empirical distribution obtained from MH or PT will be further away that the corresponding reweighed distribution. This allows us to perform a fair comparison between DCQS and MH or PT, comparing their $\ln\tilde{\mathcal{Z}}$.
Let us thus evaluate the KL distance between $\bar{\mu}(s)$ and $\mu(s)$:
\begin{align}
    \begin{split}
    \mathcal{D}(\bar{\mu}||\mu) &= \sum_{{s}\in \tilde{\mathcal{S}}} \bar{\mu}({s}) \ln\left(\frac{\bar{\mu}({s})}{\mu({s})}\right) \\
    &= \sum_{{s}\in \tilde{\mathcal{S}}} \bar{\mu}({s})\left[\ln(\bar{\mu}({s})) -  \ln({\mu}({s}))\right] \\
    &=  \ln\mathcal{Z} + \beta \ev*{E}_{\bar{\mu}} - S(\bar{\mu}).
    \end{split}
    \label{eq:kl_bar_p}
\end{align}
Here, 
\begin{equation}
    S(\bar{\mu}) \equiv -\sum_{{s}\in \tilde{\mathcal{S}}}\bar{\mu}({s})\,\ln(\bar{\mu}({s}))
\end{equation} 
is the Shannon entropy of the empirical distribution $\bar{\mu}$, and 
\begin{equation}
    \ev*{E}_{\bar{\mu}} \equiv \sum_{{s}\in \tilde{\mathcal{S}}} \bar{\mu}({s}) \,E({s})
\end{equation}
is the average energy of the empirical distribution. Using Eqs.~\eqref{eq:kl_bar_p} and~\eqref{eq:kl_tilde_p}, we see that what we want to prove, i.e. \eqlabel{eq:emp_kl},
holds if and only if
\begin{equation}
    \ln\mathcal{\tilde{Z}} \geq S(\bar{\mu}) -\beta \ev*{E}_{\bar{\mu}}.
    \label{eq:ineq_1}
\end{equation}
As we show below, this inequality is always true, and it is a manifestation of the second law of thermodynamics. Indeed, using Langrange multiplier techniques, it can be shown that the probability distribution $\bar{\mu}^*({s})$ that maximizes the entropy $S(\bar{\mu}^*)$, at fixed energy $\ev*{E}_{\bar{\mu}^*}$ and inverse temperature $\beta$, is the reweighed distribution $\tilde{\mu}({s})$, which satisfies
\begin{equation}
    S(\tilde{\mu}) = \beta \ev*{E}_{\tilde{\mu}} +  \ln\mathcal{\tilde{Z}}.
\end{equation}
Using these results, we can prove \eqlabel{eq:ineq_1}:
\begin{equation}
    S(\bar{\mu}) -\beta \ev*{E}_{\bar{\mu}} \leq S(\tilde{\mu}) -\beta \ev*{E}_{\tilde{\mu}} = \ln\mathcal{\tilde{Z}},
    \label{eq:app_2}
\end{equation}
which proves \eqlabel{eq:emp_kl}. In the first inequality in \eqlabel{eq:app_2}, we used the fact that the entropy $S(\tilde{\mu})$ of the reweighed distribution is always greater than or equal to the entropy of any other distribution with the same energy; therefore, as a particular case, it must hold for $\bar{\mu}$.

We now evaluate the statistical distance between the empirical distribution $\bar{\mu}({s})$ and the exact Boltzmann distribution $\mu({s})$. We have 
\begin{align}
    \begin{split}
    \delta(\bar{\mu},\mu) &= \frac{1}{2} \sum_{{s}\in \mathcal{S}} |\bar{\mu}({s}) -\mu({s})| \\
    &= \frac{1}{2} \sum_{{s}\in (\mathcal{S}-\tilde{\mathcal{S}})} \frac{e^{-\beta E({s})}}{\mathcal{Z}} + \frac{1}{2} \sum_{{s}\in \tilde{\mathcal{S}}} |\bar{\mu}({s}) -\mu({s})| \\
    &= \frac{1}{2} \left(1 - \frac{\tilde{\mathcal{Z}}}{\mathcal{Z}}\right) + \frac{1}{2} \sum_{{s}\in \tilde{\mathcal{S}}} |\bar{\mu}({s}) -\mu({s})|.
    \end{split}
    \label{eq:dist_bar_p}
\end{align}
We now wish to find a lower bound to the term
\begin{equation}
    \sum_{{s}\in \tilde{\mathcal{S}}} |\bar{\mu}({s}) -\mu({s})|.
\end{equation}
Using Lagrange multipliers to find the probability distribution $\bar{\mu}^*({s})$ that minimizes the previous expression, non-null only in $\tilde{\mathcal{S}}$ and satisfying the normalization condition, we find the necessary condition
\begin{equation}
    \text{sign}(\bar{\mu}^*({s}) -\mu({s})) = \text{const}
\end{equation}
for all $s\in \tilde{\mathcal{S}}$. Since $\bar{\mu}^*({s})$ sums to $1$ within $\tilde{\mathcal{S}}$, whereas $\mu({s})$ sums to something smaller or equal to one within $\tilde{\mathcal{S}}$, we see that the sign must be positive (or null). Using this result, and starting from \eqlabel{eq:dist_bar_p}, we have that
\begin{align}
    \begin{split}
\delta(\bar{\mu},\mu) &=
    \frac{1}{2} \left(1 - \frac{\tilde{\mathcal{Z}}}{\mathcal{Z}}\right) + \frac{1}{2} \sum_{{s}\in \tilde{\mathcal{S}}} |\bar{\mu}({s}) -\mu({s})| \\
    &\geq \frac{1}{2} \left(1 - \frac{\tilde{\mathcal{Z}}}{\mathcal{Z}}\right) + \frac{1}{2} \sum_{{s}\in \tilde{\mathcal{S}}} \left(\bar{\mu}^*({s}) -\frac{e^{-\beta E({s})}}{\mathcal{Z}}\right) \\ 
    &= 1 - \frac{\tilde{\mathcal{Z}}}{\mathcal{Z}} =
    \delta(\tilde{\mu},\mu).
    \end{split}
\end{align}
In the inequality, we use the fact that the sum of the absolute value can be lower-bounded by replacing $\bar{\mu}(s)$ with $\bar{\mu}^*(s)$. We then use the fact that the sign is uniform to remove the absolute value.
This concludes the proof of \eqlabel{eq:emp_tvd}.

\section{Transfer matrix approach for the 1D Ising model}
\label{app:1d_ising}
In this appendix, we provide details of the transfer matrix approach used in \seclabel{subsec:ising} of the main text. We use it to compute exact properties of the 1D Ising model, and we refer to Ref.~\cite{yeomans1992} for a detailed derivation. Let us consider the Hamiltonian in \eqlabel{eq:ising_h} of the main text, 
\begin{equation}
    \hat{H}_\text{f} = \sum_{i=1}^{N} \left[ h_i \, \hat{Z}_i + J_{i} \, \hat{Z}_{i} \hat{Z}_{i+1}  \right],
\end{equation}
with $\hat{Z}_{N+1} = \hat{Z}_1$.
Its partition function is given by
\begin{equation}
    \mathcal{Z} = \sum_{ \{s_i\}=\pm 1} \exp\left[-\beta \sum_{i=1}^{N} \left( h_i \, s_i + J_{i} \, s_{i} s_{i+1} \right)  \right],
    \label{eq:z_1d_app}
\end{equation}
with $s_{N+1} = s_1$.
To efficiently compute the partition function, we can introduce the so-called transfer matrix $T_i$ at each site, defined as 
\begin{equation}
T_i = 
\begin{pmatrix}
    e^{-\beta(J_i + \frac{h_i+h_{i+1}}{2})} &  e^{-\beta(-J_i + \frac{h_i-h_{i+1}}{2})} \\
    e^{-\beta(-J_i + \frac{-h_i+h_{i+1}}{2})} &  e^{-\beta(J_i - \frac{h_i+h_{i+1}}{2})} 
\end{pmatrix}.
\end{equation}
The sum in \eqlabel{eq:z_1d_app} can then be expressed as
\begin{equation}
    \mathcal{Z} = \Tr[ T_1T_2\dots T_N ].
\end{equation}
Since the $\{h_i\}$ and $\{J_i\}$ coefficients are non-uniform, we numerically compute the partition function by multiplying the $N$ $2\times 2$ transfer matrices and taking their trace. The time required to perform such an operation scales linearly with the number of qubits, so it is efficient to compute.

To compute the observables reported in the main text, we use the following identities, which can be readily verified using the definition of the partition function provided in \eqlabel{eq:z_1d_app}:
\begin{equation}
\begin{aligned}
    \ev*{\hat{Z}_i} &= -\frac{1}{\beta} \frac{\partial \ln\mathcal{Z}}{\partial h_i}, &
    \ev*{\hat{Z}_i\hat{Z}_{i+1}} &= -\frac{1}{\beta} \frac{\partial \ln\mathcal{Z}}{\partial J_i}, \\
    \ev*{\hat{H}_\text{f}} &= - \frac{\partial \ln\mathcal{Z}}{\partial \beta}.
\end{aligned}
\end{equation}

\begin{figure}[!tb]
    \centering
    \includegraphics[width=0.99\columnwidth]{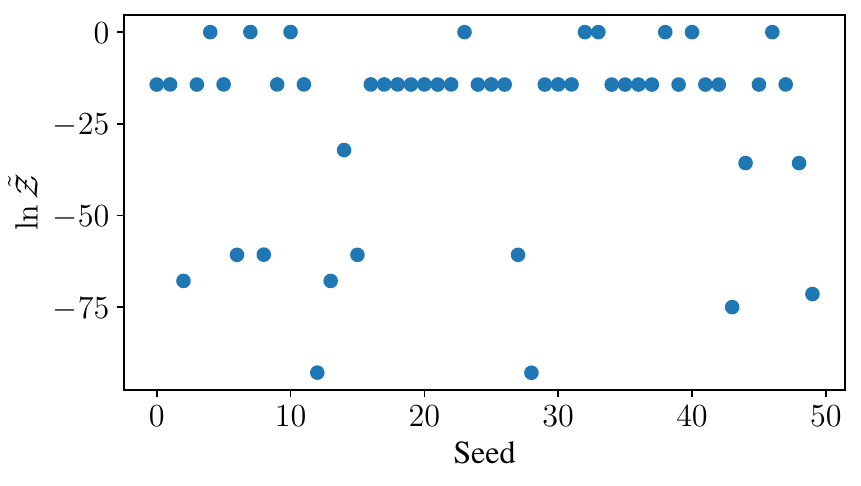}
    \caption{Value of $\ln \tilde{\mathcal{Z}}$ (up to a global offset) of DCQS at $T=0.02$, for the system studied in \seclabel{subsec:3body} of the main text, as a function of the seed for the random number generator used for the PP. Nine seeds out of $50$ reach the highest value of the low-temperature $\ln \tilde{\mathcal{Z}}$. }
    \label{fig:13}
\end{figure}

\begin{figure}[!tb]
    \centering
    \includegraphics[width=0.99\columnwidth]{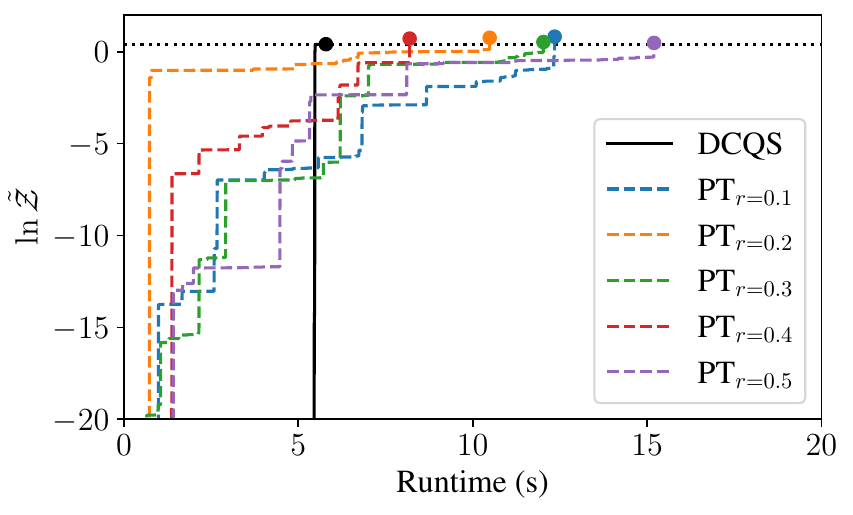}
    \caption{Same plot as \figlabel{fig:10} of the main text, but choosing a temperature $T=0.1$. DCQS is still the fastest method, although the advantage over PT is smaller compared to $T=0.02$. }
    \label{fig:14}
\end{figure}

\section{Details on DCQS for the higher order spin-glass Hamiltonian}
\label{app:sidon}

In this appendix, we provide additional details on the DCQS algorithm applied in \seclabel{subsec:3body} of the main text. For the calculation of the BF during each iteration, we proceed as follows. We perform three post-processing sweeps on the $2000$ lowest-energy states, as described in \seclabel{subsec:ising}. This is equivalent to a zero temperature MCMC simulation or a greedy energy-descent optimization. We then compute the BF as in \eqlabel{eq:bf_update}, setting $n_\text{cvar}=1$ and only considering the single lowest-energy bitstring identified during the post-processing. These states are not included in $\tilde{\mathcal{S}}$ but only temporarily used to compute the BF at each iteration. We account for the runtime necessary for the BF calculation by considering that these samples are generated at the same speed as in PT, i.e., 8.7 million samples per second.

The experiments were performed on the 156-qubit \textsc{IBM Fez} superconducting quantum processor, accessed via cloud using Qiskit~\cite{javadiabhari2024}. The DCQS circuits were transpiled using the Qiskit transpiler with optimization level~3. To further reduce circuit depth and improve compilation accuracy, we employed fractional gates~\cite{ibm_fractional_gates} that are natively supported on \textsc{IBM Heron} quantum processors. In particular, the use of $\hat{R}_{\text{zz}}(\theta)=\exp(-i\theta\hat{Z}_1\hat{Z}_2/2)$ for $0<\theta\leq\pi/2$ and $\hat{R}_\text{x}(\theta)=\exp(-i\theta\hat{X}/2)$ enabled more efficient implementation of entangling operations. Dynamical decoupling was applied throughout execution using an $X_pX_m$ pulse sequence with an \textit{as-soon-as-possible} (ASAP) schedule. This seeks to mitigate decoherence during idle periods.

To analyze the effectiveness of PP in reaching high values of the figure of merit in the low-temperature regime, in \figlabel{fig:13}, we plot the final value of $\ln \tilde{\mathcal{Z}}$ as a function of the seed used for the random number generator in PP. As we can see, $9$ seeds out of $50$ reach the largest value of $\ln \tilde{\mathcal{Z}}$. In the main text, we report the performance of DCQS using one of these $9$ seeds: seed value of $10$. However, all of these $9$ seeds lead to the same conclusions. 

\begin{figure}[!tb]
    \centering
    \includegraphics[width=0.99\columnwidth]{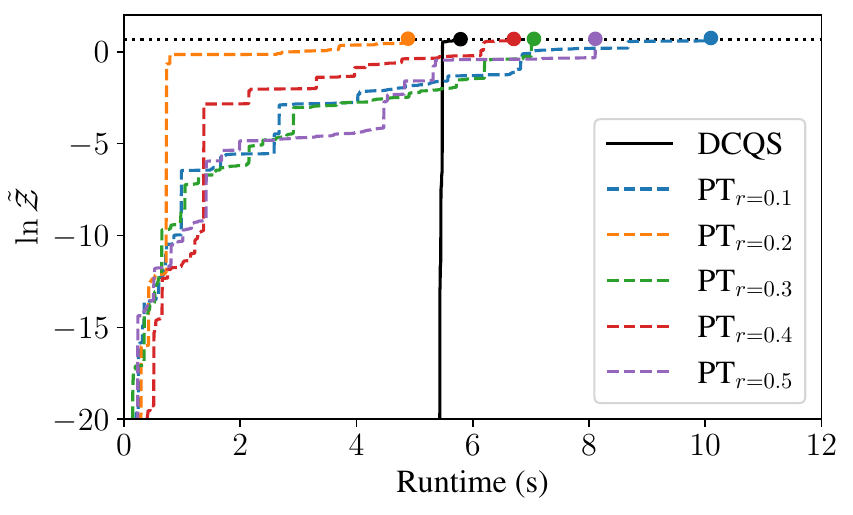}
    \caption{Same plot as \figlabel{fig:10} of the main text, but choosing a temperature $T=0.2$. DCQS is the fastest method in all cases, except for $\text{PT}_{r=0.2}$. }
    \label{fig:15}
\end{figure}
Finally, in Figs.~\ref{fig:14} and \ref{fig:15}, we reproduce \figlabel{fig:10} of the main text for the temperatures $T=0.1$ and $T=0.2$, respectively. As expected, as we increase the temperature, the advantage of DCQS over PT decreases. Indeed, in \figlabel{fig:14}, DCQS is still the fastest method, but the advantage is smaller compared to the values reported in \figlabel{fig:10} of the main text, where $T=0.02$. In \figlabel{fig:15}, however, DCQS is the second-fastest approach, with $\text{PT}_{r=0.2}$ reaching the same figure of merit with a lower runtime.

\bibliography{references}

\end{document}